# Understanding Dynamics in Coarse-Grained Models: II. Coarse-Grained Diffusion Modeled Using Hard Sphere Theory

Jaehyeok Jin,[1] Kenneth S. Schweizer,[2] and Gregory A. Voth[1*]

[1] Department of Chemistry, Chicago Center for Theoretical Chemistry, Institute for Biophysical Dynamics, and James Franck Institute, The University of Chicago, Chicago, IL 60637, USA

[2] Department of Material Science, Department of Chemistry, Department of Chemical & Biomolecular Engineering, and Materials Research Laboratory, University of Illinois, Urbana, IL 61801, USA

* Corresponding author: gavoth@uchicago.edu

**Abstract**
The first paper of this series [J. Chem. Phys. 158, 034103 (2023)] demonstrated that excess entropy scaling holds for both fine-grained and corresponding coarse-grained (CG) systems. Despite its universality, a more exact determination of the scaling relationship was not possible due to the semi-empirical nature. In this second paper, an analytical excess entropy scaling relation is derived for bottom-up CG systems. At the single-site CG resolution, effective hard sphere systems are constructed that yield near-identical dynamical properties as the target CG systems by taking advantage of how hard sphere dynamics and excess entropy can be analytically expressed in terms of the liquid packing fraction. Inspired by classical equilibrium perturbation theories and recent advances in constructing hard sphere models for predicting activated dynamics of supercooled liquids, we propose a new approach for understanding the diffusion of molecular liquids in the normal regime using hard sphere reference fluids. The proposed "fluctuation matching" is designed to have the same amplitude of long wavelength density fluctuations (dimensionless compressibility) as the CG system. Utilizing the Enskog theory to derive an expression for hard sphere diffusion coefficients, a bridge between the CG dynamics and excess entropy is then established. The CG diffusion coefficient can be roughly estimated using various equations of the state, and an accurate prediction of accelerated CG dynamics at different temperatures is also possible in advance of running any CG simulation. By introducing another layer of coarsening, these findings provide a more rigorous method to assess excess entropy scaling and understand the accelerated CG dynamics of molecular fluids.

# I. Introduction

Coarse-grained (CG) models facilitate efficient computational studies compared to conventional atomistic, or fine-grained (FG), simulations.[1-10] By averaging away the presumed unimportant degrees of freedom, CG simulations can explore much larger spatiotemporal scales of the molecular systems of interest. Among various CG approaches, bottom-up CG models are designed to approximate the many-body potential of mean force (PMF) from FG simulations. In practice, relatively short FG simulations are required to construct CG models to represent the many-dimensional PMF for the given system correctly. Even though important static (e.g., structural) correlations can be captured by employing accurate conservative interactions between the CG particles,[7, 11-14] designing a CG model with only conservative interactions may not capture dynamic properties correctly. This is due to the missing fluctuation and dissipation forces present at the FG resolution. Therefore, in Hamiltonian mechanics, CG diffusion is often accelerated compared to the reference data from atomistic simulations, and it is of great importance to systematically rationalize different diffusion behaviors between FG and CG models.

To date, a number of theories have been put forth in an attempt to understand the accelerated dynamics and to correct for these faster time scales in CG models. These efforts range from the application of the Mori-Zwanzig formalism[15-33] to time-rescaling approaches,[34-39] all for specific goals governed by different equations of motions. Readers are referred to Refs. 40, 41 for a detailed review of possibly dynamically consistent CG models and the Introduction section of the preceding paper[42] in this series (hereafter referred to as "Paper I"). To note, the Mori-Zwanzig projection operator formalism allows for a rigorous description of the CG dynamics as a form of the Generalized Langevin Equation (GLE).[43-46] Combining the time-rescaling and Mori-Zwanzig approaches, Lyubimov and Guenza developed the dynamical reconstruction approach for polymers.[47-49] In dynamical reconstruction, the acceleration factor due to coarse-graining can be analytically estimated by considering atomistic polymers as a bead-and-spring and the coarse-grained representation as a soft-colloid representation. In detail, by adopting several approximations based on polymer physics, the effective friction coefficients for both FG and CG system can be further reduced into a tractable form with an analytical acceleration factor. However, such a simplified description usually does not apply for complex atomistic systems. Namely, a practical utilization of the Mori-Zwanzig formalism is numerically challenging and seemingly intractable for general molecular systems due to the complex nature of frictional forces.[16, 50]

As an alternative to other existing approaches, in this series, we aim to leverage an apparently quasi-universal relationship in normal (not supercooled) liquids known as excess entropy scaling[51-54] to understand the accelerated dynamics in CG models and their relationship with the corresponding FG dynamics. In Paper I, we discovered that FG and CG systems follow an identical excess entropy scaling relationship with the *same* exponent for the same molecule in the FG and corresponding CG models.[42] Even though this finding uncovered a universality in the scaling relationship at the CG level with respect to its FG system, a complete understanding of the accelerated CG dynamics in relation to the FG dynamics was not possible because the excess entropy scaling relationship is intrinsically empirical and semi-quantitative.

One end goal would therefore be to more rigorously derive an exact analytical form of the scaling relationship from a systematic theory to link dynamic properties with the excess entropy. Having an analytical expression for excess entropy scaling, one can explicitly deduce the effective acceleration due to coarse-graining. Despite statistical mechanical efforts based on the mode coupling theory[55-57] and Boltzmann's formula,[58] a complete derivation of the excess entropy scaling relationship for molecular systems at the FG resolution is very challenging since FG systems exhibit complicated dynamics as different molecular motions and degrees of freedom are coupled. However, these complications are considerably reduced at the CG resolution. Especially, at the single-site CG model, where each molecule is mapped to its center-of-mass as the CG site, degrees of freedom other than translation are integrated out, resulting in single-site CG translational dynamics.[42]

In this second paper of the series, we focus solely on understanding the full CG dynamics and determining the excess entropy scaling relationship at the single-site CG resolution as the first step toward establish accurate dynamic correspondence between the FG and CG dynamics. Taking a step further from the CG viewpoint, we investigate this dynamical behavior through an even more simplified lens: the hard sphere reference fluid point of view.[59-62] Specifically, we treat the CG system as an effective hard sphere fluid. Even though the hard sphere system is characterized by an infinite repulsive interaction below the contact distance with zero attraction, an extension to molecular CG systems that have more complex interaction potentials is conceivable. Seminal works on classical equilibrium perturbation theories of liquids by Zwanzig,[63] Barker-Henderson (BH),[64, 65] and Weeks-Chandler-Andersen (WCA)[66-68] suggest that the short-range repulsive interaction determines the structure in non-associated dense liquids, while the attractive longer-range interaction of any real single-site simple fluid gives a uniformly cohesive background that effectively cancels the effect of the surrounding molecules in a dense fluid; this characteristic is known as "force cancellation."[69] Based on theoretical developments supported by computer simulations,[70, 71] hard sphere models have played an important role in describing static and dynamic properties of real dense liquids at the atomistic resolution using analytical expressions. Nevertheless, it remains unexplored where such a hard sphere treatment can be readily applied to bottom-up molecular CG models, and the case of strongly associated water is likely the most challenging system.

This study presents a comprehensive, systematic approach for representing molecular CG models via a much coarser hard sphere description. By design, a minimalist hard sphere model of CG systems is appealing due to not only the simplicity of representation at the CG resolution, but also the spherically-symmetric nature of the CG interaction, especially at a single-site resolution. Since this inherent simplicity of hard spheres allows the dynamic processes to be analytically determined, the key idea of this present work is to design a dynamically consistent hard sphere mapping theory that can reasonably estimate CG dynamics.

Classical thermodynamic and related perturbation theories serve as a theoretical basis for constructing a hard sphere mapping to predict equilibrium structure, yet conventional mapping schemes (e.g., the BH treatment) may be limited in their ability to faithfully capture dynamic properties. Indeed, the conventional approaches typically assume attractions do not change structure, and the mapping relates a hypothetical purely repulsive force reference fluid to an effective hard sphere fluid. In contrast, here we propose a mapping procedure we call "fluctuation matching" that is not tied to the precise form of the intermolecular potential of the real system of interest and which contains both repulsive and attractive interactions, but rather is based on the amplitude of long wavelength collective thermal density fluctuations that is rigorously related to a thermodynamic property. This new methodology is directly inspired by the recent theoretical work on the activated dynamics of polymer and supercooled liquids by Schweizer and coworkers.[72-77] Specifically, the dimensionless compressibility, because it is directly related to the amplitude of long wavelength density *fluctuations*, has been argued to be a natural quantity to base a mapping on for the purpose of understanding dynamic barriers and activated structural relaxation times in supercooled molecular, polymer, and other glass-forming liquids under isobaric conditions.[72-77] We are thus motivated to apply this idea to the higher temperature regime of normal (or ambient) molecular liquids at the CG level. Our focus is to build a bridge between CG dynamics and excess entropy scaling in a dynamical regime not dominated by strongly activated processes.

Based on a hard sphere system described by a packing fraction or effective hard sphere diameter (EHSD), we derive here an analytical expression of the diffusion coefficient using elementary kinetic theory. The two main theoretical findings are described in Section II. Combined with the excess entropy scaling formalism, we arrive at an analytical form of the "entropy-free" diffusion coefficient for the CG resolution. We evaluate this quantity by approximating hard sphere systems using various well-known equations of state (EOS). Comparisons to the reference diffusion coefficient obtained from molecular simulations in Paper I are presented in Section III. Section IV concludes the present Paper II and lays the foundation to resolve unaddressed questions for the next paper of the series.

## II. Theory
### A. Excess Entropy Scaling for CG Systems

We introduced the concept of excess entropy scaling and discussed its universality in molecular systems upon the coarse-graining process in Paper I, and we briefly summarize the important findings here. The excess entropy is defined as the entropy difference between the system and its corresponding ideal gas under the same temperature and density conditions

$$S_{ex} = S_{ex}(\rho, T) \coloneqq S(\rho, T) - S_{id}(\rho, T).$$

(1)

In order to quantify the quasi-empirical relations between dynamic properties and excess entropy, two different scaling schemes have been proposed in the literature. The first scheme that links the dynamic property of the self-diffusion coefficient $D$ to the molar excess entropy $s_{ex} = S_{ex}/Nk_B$ was suggested by Rosenfeld:[51-53]

$$D^* = D_0 \exp(\alpha s_{ex}),$$

(2)

where the reduced diffusion coefficient is scaled by "macroscopic" quantities (involving density, temperature, mass):

$$D^* = D \frac{\rho^{\frac{1}{3}}}{\left(\frac{k_B T}{m}\right)^{\frac{1}{2}}}.$$

(3)

On the other hand, an alternative scaling scheme has been suggested using a microscopic description by Dzugutov:[54]

$$D_Z^* = D_Z^0 \exp(s_{ex}^{(2)}),$$
(4)

where $s_{ex}^{(2)}$ is the molar pair excess entropy and is different from the full excess entropy $s_{ex}$ in Eq. (2). In the Dzugutov scaling, the diffusion coefficient is reduced based on dynamic caging:

$$D_Z^* = \frac{D}{\sigma^2 \Gamma_E} = \frac{D}{\sigma^2} \cdot \frac{1}{4\pi\sigma^2 \rho g(\sigma) \sqrt{\frac{k_B T}{\pi m}}},$$
(5)

where $\sigma$ denotes a contact distance, and $\Gamma_E = 4\pi\sigma^2 \rho g(\sigma)\sqrt{k_B T/\pi m}$ is an Enskog collision frequency. Equation (5) uses the hard sphere Enskog frequency to scale the diffusion coefficient, as Eq. (4) is derived from the hard sphere system, which certainly is not as realistic as the true molecular system. Due to this assumption, we discovered that the Rosenfeld scaling imparts a more general relationship, thus making it more relevant to linking the FG and CG systems. Comparisons between the two scaling schemes can be found in Paper I.[42]

Using Eq. (2), in Paper I,[42] we discovered that FG water (SPC/E, SPC/Fw, TIP4P/2005, and TIP4P/Ice) and the CG water model (see Section II-G for detail) follow the Rosenfeld relationship with near-identical slopes, i.e., $D_{FG}^* = D_0^{FG} \exp(\alpha^{FG} s_{ex}^{FG})$ and $D_{CG}^* = D_0^{CG} \exp(\alpha^{CG} s_{ex}^{CG})$, with respect to the excess entropy:

$$\ln D_{FG}^* = \alpha^{FG} s_{ex}^{FG} + \ln D_0^{FG} = 0.73 \times s_{ex}^{FG} + 2.15,$$
(6)

$$\ln D_{CG}^* = \alpha^{CG} s_{ex}^{CG} + \ln D_0^{CG} = 0.7 \times s_{ex}^{CG} - 0.35.$$
(7)

We note that Eqs. (6) and (7) hold regardless of the choices of force fields for water at different temperatures and densities, indicating the invariant nature of the scaling relationships under the same molecular condition. Since differences in the $s_{ex}$ terms can be explained by information loss,[78] i.e., missing configurational entropies upon coarse-graining,[79] elucidating the $D_0$ variable and its underlying difference between the FG and CG systems can help to bridge between the FG and CG descriptions of center-of-mass diffusion. Due to the complexities at the FG resolution and an absence of rigorous theory for deriving the excess entropy scaling, our primary goal here is instead to understand the physical meaning of $D_0^{CG}$, which is hereafter denoted as the "entropy-free" diffusion coefficient at the CG resolution: $\ln D_0^{CG} = -0.35$ for CG water.

**B. Hard Sphere Mapping: General Framework**

An analytical expression of $D_0$ for the given CG system is obtained through three steps. Figure 1 illustrates a schematic diagram of the hard sphere mapping procedure.

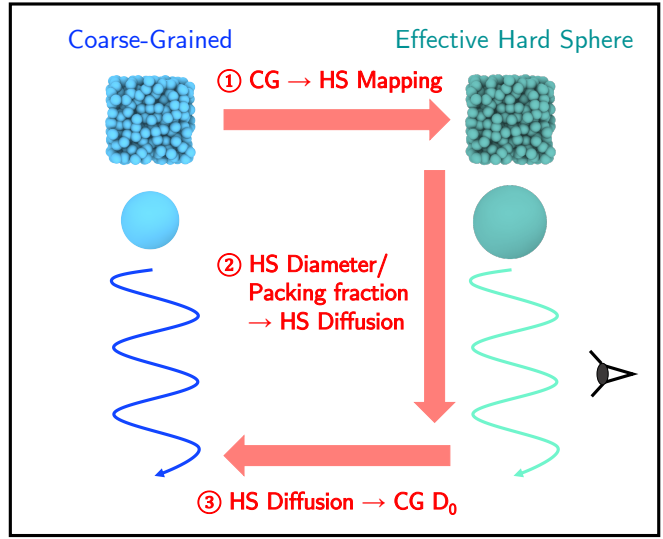

**Figure 1.** Schematic diagram describing the overall procedure of this work: (1) Map a given CG system to an effective hard sphere system. This step gives an EHSD or packing fraction of the CG model. (2) From the EHSD or packing fraction, derive an analytical expression for the diffusion coefficient of the hard sphere system. Since the hard sphere diffusion process can be correctly described by the Dzugutov scaling if the packing fraction is not too high, this step is done on the basis of the Dzugutov relationship. (3) From the hard sphere diffusion coefficient expression, infer the $D_0$ term in the Rosenfeld scaling of the original CG system.

We first map the CG system to the corresponding hard sphere system. Since dimensionless static or dynamic properties of literal hard spheres are completely described by their packing fraction $\eta$, the first step is to determine values of $\eta$ that can reproduce the important properties of the reference CG system. Conventionally, one assumes the number density is known, and this involves determining the EHSD, $\sigma$. Then, $\eta$ is readily obtained from the EHSD by the following relationship:

$$\eta = \frac{\pi}{6}\sigma^3 \rho = \frac{\pi}{6}\sigma^3 \frac{N_{CG}}{V},$$

(8)

where the system volume $V$ is identical to that of the CG system composed of $N_{CG}$ particles. However, determining $\sigma$ may not be the only way to construct a dynamically consistent effective hard sphere fluid. In Section III, we introduce a simpler approach that allows for directly mapping $\eta$.

Once the hard sphere packing fraction is determined from a mapping, the diffusion coefficient of the specific hard sphere system can be straightforwardly calculated using the hard sphere kinetic theory. Since we are dealing with the mapped hard sphere system, we first revisit the Dzugutov scaling scheme. Unlike generally complex FG systems, the Dzugutov scaling is valid only for the hard sphere systems. Thus, our ultimate goal is to analytically derive the $D_0$ term for the original Rosenfeld scaling scheme by introducing what we perceive as a dynamically consistent hard sphere reference as a proxy, where its fundamental quantities can be analytically derived using the Dzugutov scaling. Altogether, we build a dynamical correspondence between the CG and hard sphere fluids (Step 1 in Fig. 1) in order to analytically derive the $D_0$ expression in the hard sphere systems (Step 2 in Fig. 1) and transfer $D_0$ back to the original CG system (Step 3 in Fig. 1).

To calculate the static and dynamic properties of a hard sphere fluid, one must carefully choose the proper EOS. By definition, an EOS is an analytical relationship among pressure, volume, and temperature, often given in terms of the compressibility factor $\mathbb{Z}$:

$$\mathbb{Z} = \frac{P}{\rho k_B T}, \tag{9}$$

where $\mathbb{Z}$ is a function of $\eta$. To accurately represent $P(\rho, T)$ over a range of $\rho$ and $T$, an EOS is often obtained by accurately fitting to the computed values of the virial EOS:

$$\mathbb{Z}(\rho) = 1 + B_2\rho + B_3\rho^2 + \cdots = \left(1 + \sum_{k=2}^{\infty} B_k \rho^{k-1}\right), \tag{10}$$

or in terms of $\eta$ as

$$\mathbb{Z}(\eta) = 1 + 2^2\eta + b_3\eta^2 + b_4\eta^3 + \cdots = 1 + 4\eta + \sum_{k'=3}^{\infty} b_{k'}\eta^{k'-1}, \tag{11}$$

where the virial coefficient is rescaled as $b_k \coloneqq \left(\frac{\pi}{6}\sigma^3\right)^{1-k} B_k$ in three dimensions. The particular choice of the analytic form for the EOS is vital for applying the perturbation theory and calculating various thermodynamic quantities of interest,[80-83] even for systems exhibiting non-negligible attractive interactions. While classical liquid state theory asserts that such a hard sphere description cannot be directly applied at the microscopic level to associated liquids characterized by strong specific attractions (e.g., water), the goal of this paper is to demonstrate how this minimalist description can still work rather well for molecular CG models based on the mapping formulated for the problem of supercooled molecular liquid relaxation.[72-77] A detailed discussion of this practical extension will be given in Section II-E.

### C. Conventional Hard Sphere Mapping: Barker-Henderson Criterion
Classical perturbation theories of atomic or molecular (at the interaction site level) liquids treat pair interactions by separating the short-range repulsions as a hard sphere reference and the long-range interactions as a perturbation (often attractions), assuming that the structure (and sometimes the dynamics) of the liquid is determined by repulsive interactions. With this in mind, for a system with soft-core repulsive interactions as relevant to our CG model, an accurate estimation of the EHSD is needed to calculate the static and dynamic properties of the corresponding hard sphere model. In this light, various free energy perturbation theories have been developed based on different criteria to determine the EHSD. To name a few, BH[64, 65] and WCA[66-68] theories are two frequently used approaches.

Barker and Henderson proposed the first perturbation theory for simple Lennard-Jones systems. In brief, the short-range repulsive potential [positive part of the overall potential $U(R)$] was mapped to a hard sphere reference, and attractive long-range terms were treated as a perturbation.[64,65] At leading order, the simplest BH EHSD was then defined as

$$\sigma_{BH} = \int_0^{R_0} [1 - \exp(-\beta U(R))] \cdot dR,$$

(12)

where $R_0$ is the distance where the interaction vanishes, i.e., $U(R_0) = 0$. In this work, we use the many-body CG PMF from our prior work to determine the BH EHSD, which is different from the original work, where it corresponds to the Lennard-Jones interaction $U_{LJ}(R)$.

**D. Weeks-Chandler-Andersen Criterion and its Drawbacks**
The well-known WCA perturbation theory starts from an alternative, physically motivated separation method based on the sign of forces (positive or negative) rather than interaction energies.[66-68] In doing so, the WCA perturbation theory provides a smoothly varying first-order perturbative treatment of free energies. The WCA EHSD criterion is derived by approximating the cavity correlation function of the reference system $y_{HS}(R) := g(R) \exp(\beta U(R))$ as that of the hard sphere system, i.e., equating the long wavelength responses in Fourier space between the hard sphere and the purely repulsive force from reference system.[68] Mathematically, this requirement is provided by the static structure factor in a Fourier space

$$S(k) = \frac{1}{N} \langle \rho(\vec{k}) \rho(-\vec{k}) \rangle,$$

(13)

where $\rho(\vec{k}) = \sum_j^N \exp(i\vec{k} \cdot \vec{r}_j)$. Alternatively, using the radial distribution function (RDF), Eq. (13) can be rewritten as

$$S(k) = 1 + 4\pi\rho \int_0^\infty dR \, R^2 \frac{\sin(kR)}{kR} [g(R) - 1].$$

(14)

In the $k = 0$ limit, the structure factor reduces to

$$S(k = 0) = 1 + 4\pi\rho \int_0^\infty dr \, R^2 [g(R) - 1].$$

(15)

Therefore, based on the WCA separation of the pair potential, the EHSD of the effective hard sphere fluid that best captures the behavior of the purely repulsive fluid is determined by matching the $S(k = 0)$ (dimensionless compressibility) quantities:

$$\int_0^\infty y_{HS}(R; \sigma_{WCA}) \{\exp[-\beta U_{WCA}(R)] - \exp[-\beta U_{HS}(R; \sigma_{WCA})]\} R^2 \cdot dR = 0.$$

(16)

In practice, this procedure is conducted iteratively until $\sigma_{WCA}$ converges. It has been shown that the WCA theory is superior to the BH description for atomic liquids, especially at the high densities of present interest.[59-62]

However, the WCA theory suffers from three drawbacks pertinent to this study. First, a determination of $\sigma_{WCA}$ requires solving Eq. (16) iteratively, making it less efficient. Also, in our cases, for both the BH and WCA theories, the EHSD is estimated by many-dimensional CG PMFs projected on pairwise basis sets. Furthermore, it is worthwhile noting that the original BH and WCA approaches were designed for systems with harsh repulsive interactions, and extending these approaches to systems with soft repulsions, in which attractions vary rapidly in space and can change structure, is of unknown reliability. Given the complex interaction profile of CG PMFs of molecular liquids, we envisage that the EHSD scheme based on $U(R)$ might be less accurate when applied to CG systems. This suggests that such a method should focus on determining the effective packing fraction for which the density and hard sphere diameter are not separable. Lastly, both the BH and WCA criteria for the EHSD are derived from the perspective of describing equilibrium thermodynamic properties (e.g., free energy), and thus it is not *a priori* clear how good they are to capture CG dynamics. In this regard, a mapping method that matches a quantity believed to be strongly correlated with dynamics is desirable. We note that the dynamical reconstruction approach also employs a reference hard sphere model to understand the CG diffusion of polymers.[47-49] However, in Refs. 47-49, the EHSD is determined via the long-time limit of Rouse dynamics of polymer melts, and hence the underlying physical implementation of the hard sphere description are different in both cases.

### E. Alternative Hard Sphere Mapping: Fluctuation Matching

To reconcile the aforementioned issues, we employ a hard sphere mapping method that has been highly developed and applied by Schweizer and coworkers.[72-77] It is dynamically motivated and has been shown to be useful for mapping real molecular and polymer liquids with both repulsive and attractive interactions to an effective hard sphere description for use in a dynamical theory of activated relaxation.[72-77] The central idea is to require the dimensionless compressibility, $S(k = 0)$, of Eq. (17) be exactly equal to its value for the real thermal liquid at all temperatures and pressures. Note that the right-hand side of Eq. (15) is related[59] to the isothermal compressibility, defined as $\kappa_T = -\frac{1}{V}\left(\frac{\partial V}{\partial P}\right)_T$, and can be rewritten as

$$S(k = 0) = \rho k_B T \kappa_T. \tag{17}$$

This quantity also enters WCA theory, but there it is employed to only match the dimensionless amplitude of long wavelength density fluctuations of the purely repulsive reference fluid and its effective hard sphere analog for atomic liquids. The criterion of matching of $S(k = 0)$ of attractive thermal liquids and a reference hard sphere fluid yields an effective hard sphere packing fraction that is chemistry and thermodynamic state dependent, and there is no need to explicitly know the effective hard sphere diameter. Under isobaric conditions, attractions can strongly modify density as the liquid is cooled, and this important effect enters the mapping procedure. The idea to use such a $S(k = 0)$ mapping approach in a dynamical context was motivated by the physical argument that collective density fluctuations is the key slow variable that quantifies caging effects in dense liquids.[72-77] When this mapping is combined with the activated dynamics theories of Refs. 72-77, an understanding of the dependence of the structural relaxation time of chemically complex molecular and polymer liquids over many orders of magnitudes, extending from the weakly to strongly supercooled regimes down to the kinetic glass transition, has been achieved.[72-77] We believe this provides a firm foundation for our present work in the normal liquid regime where the dynamical problem is, in fact, simpler than in the supercooled regime since caging effects are far weaker.

Here, we extend the above ideas to CG models of molecular liquids in the normal state regime. *A priori*, the accuracy of this approach is unknown given the inter-particle potentials are softer and can have strong attractions, especially for water. The use of this mapping assumes these features do not change the usefulness of the basic idea that the effect of dynamic correlations and caging remains strongly correlated with collective density fluctuation dynamics. For hard spheres, since the isothermal compressibility is given by the derivative of the EOS from the compressibility route, one only needs the effective packing fraction, $\eta$, i.e., $S(k \to 0)_{HS} = f(\eta)$.

In the present context, we call this approach "fluctuation matching", which is defined mathematically as equating the dimensionless long wavelength density fluctuation amplitude of the CG and hard sphere systems:

$$S(k = 0)_{CG} = S(k = 0)_{HS}. \tag{18}$$

We emphasize this mapping via $S(k = 0)$ directly focuses on $\eta$, and hence avoids the issue of determining the effective hard sphere diameter with $\eta$ carrying the chemistry, temperature, and pressure dependences. This distinction enhances the power of a hard sphere mapping in terms of transferability, and employing only $\eta$ rather than separating it into EHSD is consistent with Rosenfeld's original work.[51-53] In practice, for a spherical particle system (i.e., a single site CG model), the right-hand side of Eq. (18) reduces to a direct connection between the dimensionless compressibility and the RDF of any spherical particle fluid:

$$f(\eta) = 1 + 4\pi\rho \int_0^\infty dr\, R^2[g(R) - 1]. \tag{19}$$

Of course, *a priori*, one might expect that this proposed approach will face a significant challenge in the application to molecular CG systems that exhibit liquid dynamics influenced by strong- and short-range attractive interactions (associated liquids). While recent work in this direction has provided a deeper level of understanding of the connections between $S(k = 0)$ and local structure,[84-88] being able to match dimensionless compressibilities is not guaranteed to be an accurate strategy in a dynamical context for associated normal state liquids. Moreover, the adoption of a hard sphere reference system and the above mapping effectively assumes that although attractions do modify the magnitude, temperature and pressure dependence of the effective packing fraction, and hence structure, for dynamics it is assumed repulsive forces dominate. References 72-77, where the analog of $S(k = 0)_{CG}$ is the experimental dimensionless compressibility under isobaric conditions, have shown success for supercooled (largely non-associated) molecular and polymer liquids and provides support for this simplification. Concerning practical implementation, as one sees from Eq. (19) the key quantity $f(\eta)$ is dependent on $\kappa_T$ which is determined from the EOS. In later sections, we will determine the full form of Eq. (18) using several hard sphere fluid EOSs along with the FG simulations. This will be a main focus of the rest of this paper.

## F. Hard Sphere Diffusion Coefficient from Hard Sphere Diameter
### 1. Excess Entropy Scaling

The previous section demonstrated that CG systems can be mapped onto effective hard sphere fluids by utilizing either the BH or fluctuation matching. With this in mind, we now derive an analytical expression of $D_0^{CG}$ for the Rosenfeld scaling in terms of effective hard sphere packing fraction.

Our starting point is the alternative Dzugutov scaling relationship that accurately holds for the hard sphere fluid as well as for the Rosenfeld scaling: $D_Z^* = D \cdot (\sigma^{-2}\Gamma_E^{-1}) = D_Z^0 \exp(s_{ex}^{(2)})$. For the hard sphere system, we determine $s_{ex}^{(2)}$ from the excess entropy of the hard sphere fluid, $s_{ex}^{HS}$, that can be computed analytically in many cases.[89, 90] We thus write the unscaled diffusion coefficient $D$ of the hard sphere system as

$$D = \sigma^2 \Gamma_E D_Z^0 \exp(s_{ex}^{HS}).$$
(20)

An advantage of the hard sphere system is that the hard sphere diffusion coefficient in Eq. (20) can be alternatively expressed based on Enskog kinetic theory.[91, 92] By considering the positional correlations in a fluid of spheres, Enskog theory has been shown to provide a reasonably good description of dynamics up to moderate liquid-like densities.[93-95] We adopt the Enskog expression for the hard sphere diffusion coefficient:

$$D = D^{HS} = \frac{3}{8\pi}\sqrt{\frac{\pi k_B T}{m}}\frac{1}{\rho\sigma^2 g(\sigma)}.$$
(21)

Equating Eq. (20) with Eq. (21) yields

$$D_Z^0 = \frac{3}{32\pi}\frac{\exp(-S_{ex}^{HS})}{\rho^2 \sigma^6 g^2(\sigma)}.$$
(22)

An explicit form of Eq. (22) can be fully determined by the EOS via the compressibility factor $\mathbb{Z}$, which from the thermodynamic virial route is given by

$$g(\sigma) = \frac{\mathbb{Z}-1}{4\eta},$$
(23)

where the RDF at contact, $g(\sigma)$, is directly related to the reduced pressure. Furthermore, the excess entropy of the hard sphere system can be obtained from

$$s_{ex}^{HS} = -\int_0^\eta \frac{\mathbb{Z}-1}{\eta'}d\eta'.$$
(24)

Since $\mathbb{Z}$ is a function of the packing fraction $\eta$, $D_Z^0$ is entirely determined from $\eta$ as follows:

$$D_Z^0 = \frac{\pi}{24} \cdot \frac{1}{(\mathbb{Z}-1)^2} \exp\left(\int_0^\eta \frac{\mathbb{Z}-1}{\eta'} d\eta'\right).$$

(25)

**2. Equation of State**

We now derive an analytic expression for $D_0$ using the analytically solvable Percus-Yevick (PY) integral equation theory.[96] However, PY theory is thermodynamic inconsistent,[97] meaning that thermodynamic quantities derived from identical RDFs by different formally exact routes are not identical. The compressibility factor $\mathbb{Z} := \frac{P}{\rho k_B T}$ derived from the PY theory via the compressibility route is

$$\mathbb{Z}_{PY}^c = \frac{(1+\eta+\eta^2)}{(1-\eta)^3},$$

(26)

which agrees with the Scaled Particle Theory.[98-100] However, an alternative $\mathbb{Z}$ expression derived by Wertheim and Thiele from the virial route is[101-103]

$$\mathbb{Z}_{PY}^v = \frac{(1+2\eta+3\eta^2)}{(1-\eta)^2}.$$

(27)

In this work, we chose the compressibility route $\mathbb{Z}_{PY}^c$, since overall $\mathbb{Z}_{PY}^c$ provides better description of virial coefficients[104] and compressibility factors[105] than $\mathbb{Z}_{PY}^v$, and because the fluctuation matching mapping strategy focuses on the dimensionless compressibility. In the Appendix we derive the relevant static and dynamic quantities using the virial route.

Using the EOS as Eq. (26), the excess entropy predicted from the PY EOS is:

$$s_{ex}^{HS} = -\int_0^\eta \frac{\eta'^2 - 2\eta + 4}{(1-\eta')^3} d\eta' = \ln(1-\eta) - \frac{3}{2}\left[\frac{1}{(1-\eta)^2} - 1\right],$$

(28)

and the pair correlation function at contact $g(\sigma)$ is

$$g(\sigma) = \frac{\eta^2 - 2\eta + 4}{4(1-\eta)^3}.$$

(29)

**3. Final Expression for $D_0$**

Combining Eqs. (28) and (29), we arrive at the $D_Z^0$ expression based on the PY EOS (compressibility route), which we denote as $D_{Z,PY}^0$:

$$D_{Z,PY}^0 = \frac{\pi}{24} \cdot \frac{(1-\eta)^5}{\eta^2(\eta^2 - 2\eta + 4)^2} \exp\left[\frac{3(2\eta - \eta^2)}{2(1-\eta)^2}\right].$$

(30)

Therefore, the diffusion coefficient of the hard sphere using the Dzugutov scaling $D = \sigma^2 \Gamma_E D_Z^0 \exp(S_{ex})$ is

$$D = 4\pi\sigma^4 \rho\, g(\sigma) \left(\frac{k_B T}{\pi m}\right)^{\frac{1}{2}} \times \frac{\pi}{24} \cdot \frac{(1-\eta)^5}{\eta^2(\eta^2 - 2\eta + 4)^2} \exp\left[\frac{3(2\eta - \eta^2)}{2(1-\eta)^2}\right] \exp(s_{ex}).$$
(31)

Now, writing $\frac{\pi}{24}\sigma^4\rho = \frac{\sigma}{4} \cdot \left(\frac{\pi}{6}\sigma^3\rho\right) = \frac{\sigma}{4} \cdot \eta$, we obtain

$$D = \sigma^2 \Gamma_E D_Z^0 \exp(s_{ex}) = \frac{\sigma}{4}\left(\frac{\pi k_B T}{m}\right)^{\frac{1}{2}} \times \frac{(1-\eta)^2}{\eta(\eta^2 - 2\eta + 4)} \exp\left[\frac{3(2\eta - \eta^2)}{2(1-\eta)^2}\right] \exp(s_{ex}).$$
(32)

We now seek to transfer $D$ (Step 3 in Fig. 1) back to the $D_0$ of the CG system via the Rosenfeld scaling. From the "macroscopic" rescaling in Eq. (3), we arrive at

$$D^* = D \cdot \frac{\rho^{\frac{1}{3}}}{\left(\frac{k_B T}{m}\right)^{\frac{1}{2}}} = \frac{6^{\frac{1}{3}}}{4}\pi^{\frac{1}{6}}\eta^{\frac{1}{3}} \frac{(1-\eta)^2}{\eta(\eta^2 - 2\eta + 4)} \exp\left[\frac{3(2\eta - \eta^2)}{2(1-\eta)^2}\right] \exp(s_{ex}).$$
(33)

An empirical aspect of the Rosenfeld scaling originates from an exponent $\alpha$, which hinders the derivation of a general and analytic expression. However, since the excess entropy of the hard sphere fluid should be always larger (more positive) than the molecular CG system due to missing degrees of freedom, the origin of $\alpha$ arises from $\exp(\alpha^{CG} s_{ex}^{CG}) / \exp(s_{ex}^{HS}) \approx 1$. We confirmed that this *back-mapping approximation* holds for the molecular system studied in this work. Detailed analysis and the underlying physical principles of this approximation will be pursued in a future article.

Finally, the Rosenfeld scaling is applied to Eq. (33), giving:

$$D_0^{PY} \approx \frac{6^{\frac{1}{3}}}{4}\pi^{\frac{1}{6}}\eta^{\frac{1}{3}} \frac{(1-\eta)^2}{\eta(\eta^2 - 2\eta + 4)} \exp\left[\frac{3(2\eta - \eta^2)}{2(1-\eta)^2}\right].$$
(34)

Equation (34) again highlights our central idea that the diffusion coefficient in a single-site CG model is solely governed by the packing fraction of the effective hard sphere system.

To summarize, in this subsection, we derived an approximate expression for $D_0$ for molecular CG systems. While analytically deriving such an expression is not practically feasible for the real CG systems, we overcame this bottleneck by representing the fluid of CG particles as effective hard spheres. Since the dynamics of hard spheres can be (approximately) described analytically by both kinetic theory and the alternative scaling relationship by Dzugutov, we showed that for a given equation of state, $D_0$ can be expressed as a function of the packing fraction if one could faithfully map the molecular CG system to its corresponding hard sphere system.

We note that the present approach is somewhat similar to the previous work of Bretonnet, where the semi-empirical relationship between the hard sphere diffusion coefficient and the Dzugutov scaling was discussed.[89] Furthermore, Bretonnet derived an analytical expression using only the PY EOS as determined from the virial route, $\mathbb{Z}_{PY}^v$, for Dzugutov scaling. Since the Dzugutov relationship is limited to hard spheres, the novelty of the current work is to extend the Enskog theory to an effective CG system under a more general scaling relationship with an improved hard sphere EOS [Eq. (34)]. Results for other choices of the EOS are discussed in the Results section below.

**G. Computational Models**

In this paper series, we utilize our recently developed CG model for water: Bottom-Up Many-Body Water (BUMPer).[106, 107] In Paper I, we have demonstrated the excess entropy scaling relationship of the BUMPer model parametrized by the SPC/E, SPC/Fw, TIP4P/2005, and TIP4P/Ice force fields follows Eq. (7).[42] In principle, any other CG model for water, e.g., monatomic Water (mW),[108] can also be applied for this framework. While the strong orientational preferences of water may require higher-order interactions beyond the two-body interaction,[109, 110] the BUMPer model is designed to reproduce many-body correlations using only pairwise basis sets. This is because the force-matched three-body interactions are in the form of the Stillinger-Weber three-body interaction[111] and can be effectively integrated into the short-range pairwise interactions of BUMPer. In this regard, BUMPer is computationally inexpensive compared to atomistic simulations, or even compared to other CG models with explicit three-body interactions. Despite having only pairwise interactions, our recent findings also demonstrated that BUMPer can accurately capture both pairwise and many-body density correlations.[106] The explicitly integrated three-body contributions in BUMPer are further corroborated by its ability to capture nucleation at the ice/water interface.[107] Moreover, a pairwise decomposable form of the BUMPer interaction allows for the excess entropy scaling to be valid, unlike other models with explicit many-body interactions.[112-114]

One of the central assumptions made in this work originates from the classical perturbation theory asserting that, for a dense liquid, repulsive forces dominate the liquid structure. However, it is known that this idea cannot generally be applied for highly polar liquids, ionic solutions, and water.[115] This is because strong attractive interactions, such as hydrogen bonding for water at the atomistic resolution, rapidly vary as an interparticle distance changes, and thus they play a non-negligible role in the structure.[116] Nevertheless, we note that the current scope of this work is to apply a minimalist hard sphere mapping to water at the *CG* resolution. In contrast to the atomistic resolution, hydrogen bonding interactions do *not* directly appear at the current CG level, as they are implicitly folded into the CG model. Thus, we believe it is sensible to explore the possible usefulness of an effective hard sphere mapping scheme for the CG water system without any strong interactions to account for hydrogen bonding as in the sticky hard sphere model.[117]

Below we analyze dynamical properties (diffusion coefficient) using the BUMPer model parameterized from four different atomistic force fields at 300 K conditions: SPC/E, SPC/Fw, TIP4P/2005, and TIP4P/ice. The procedure to compute the dynamical properties is discussed in Paper I,[42] and the performance of the BUMPer models in terms of structural correlations is extensively analyzed in the original BUMPer paper series.[106, 107]

## III. Results

A primary goal of this section is to apply the aforementioned theories for molecular CG systems in order to derive the $D_0$ values. The organization of this section follows that of Fig. 1. In the first step, we discuss two different approaches for obtaining the effective hard sphere diameter or packing fraction for the given CG systems. In the second step, we utilize Eqs. (20)-(34) to estimate the $D_0$ values for various equations of state. The accuracy of the present approach is evaluated by comparing it to the actual $D_0$ values obtained from the scaling relationship.

### A. Step 1: Barker-Henderson Approach

The BH EHSD is directly determined from the many-body CG PMFs. Therefore, compared to the fluctuation matching approach, the BH EHSD is not affected by EOS choices. Figure 2 illustrates the many-body CG PMF used in this work from Paper I[42] and Refs. 106, 107.

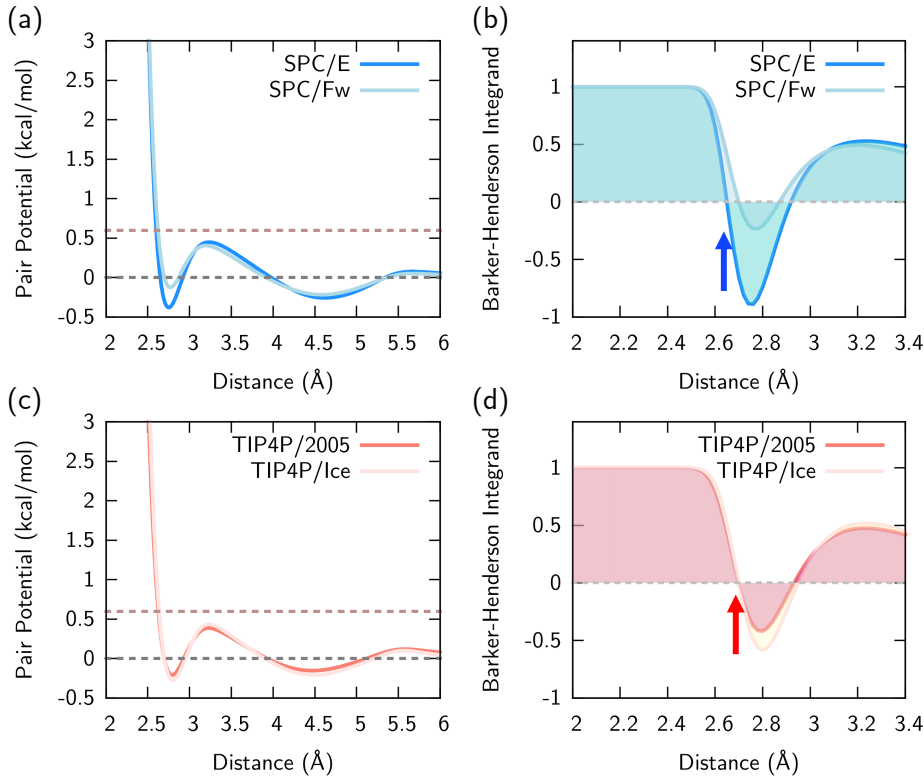

**Figure 2.** Many-body CG PMFs and corresponding effective hard sphere diameters via the BH perturbation theory. Effective CG pair potentials of the BUMPer CG models that are parameterized from (a) SPC-based force fields (SPC/E: blue, SPC/Fw: sky-blue) and (c) TIP4P-based force fields (TIP4P/2005: red, TIP4P/Ice: ivory). From each CG pair potential, we calculate the BH diameter by integrating the integrand $1 - \exp(-\beta U(R))$ over the restricted range of separations shown in (b) and (d).

The four BUMPer interactions studied in this work are parameterized from FG simulations using SPC-type (Fig. 2(a)) and TIP4P-type (Fig. 2(c)) force fields. Regardless of FG force fields, the resultant BUMPer interactions have relatively similar profiles, represented by two characteristic length scales corresponding to hydrogen bonding and van der Waals interactions. From the pairwise CG interaction, the BH EHSD was calculated by employing $\int_0^{R_0}[1 - \exp(-\beta U(R))] \cdot$ $dR$ where the BH integrands are depicted in Fig. 2(b) and 2(d), respectively. Here, we chose $R_0$ as the shortest pair distance with a zero potential value, resulting in EHSD values of 2.549 Å for SPC/Fw, 2.527 Å for SPC/E, 2.563 Å for TIP4P/2005, and 2.552 Å for TIP4P/Ice.

**B. Step 1: Fluctuation Matching**
**1. Percus-Yevick EOS**
By equating Eq. (18) to Eq. (19), the fluctuation matching approach obtains $\eta$ by solving the following equation

$$f(\eta)|_{\text{HS}} = 1 + 4\pi\rho \int_0^\infty dr\, R^2[g(R) - 1]\Big|_{\text{CG}}.$$
(35)

Note that the right-hand side of Eq. (35) is evaluated from the CG simulation to fit $f(\eta)$ from the hard sphere reference system. To determine $f(\eta)$ analytically, recall that the PY EOS via the compressibility route is given as

$$\mathbb{Z}_{\text{PY}} = \frac{(1 + \eta + \eta^2)}{(1 - \eta)^3}.$$
(36)

Then, the isothermal compressibility $\kappa_T$ of the PY EOS can be obtained using the chain rule

$$\left(\frac{\partial P}{\partial V}\right) = \left(\frac{\partial P}{\partial \eta}\right) \cdot \left(\frac{\partial \eta}{\partial V}\right) = -\left(\frac{\partial P}{\partial \eta}\right) \cdot \frac{\eta}{V}.$$
(37)

The last term $\eta/V$ is from the definition of the packing fraction: $\eta = \frac{\pi}{6}\sigma^3 \frac{N}{V}$. Further simplification is possible using the compressibility factor $\mathbb{Z}_{\text{PY}}$

$$\left(\frac{\partial P}{\partial V}\right) = -\frac{\partial}{\partial \eta}\left(\mathbb{Z}_{\text{PY}} \cdot \frac{N\beta}{V}\right) \cdot \frac{\eta}{V} = -\frac{6}{\pi\sigma^3} \cdot \frac{\partial}{\partial \eta}(\eta \mathbb{Z}_{\text{PY}}) \cdot \frac{\eta}{V}.$$
(38)

Substituting Eq. (36) into Eq. (38) gives

$$\left(\frac{\partial P}{\partial V}\right) = -\frac{\rho}{\beta V} \cdot \frac{(1 + 2\eta)^2}{(1 - \eta)^4}.$$
(39)

Finally, we arrive at the well-known dimensionless compressibility relationship of hard spheres using the PY EOS

$$S(k = 0)_{\text{HS}}^{\text{PY}} = \rho k_B T \cdot \left(-\frac{1}{V} \cdot \frac{\partial V}{\partial P}\right) = \frac{(1 - \eta)^4}{(1 + 2\eta)^2}.$$
(40)

Analytically, Equation (40) has four solutions, but only one solution $\eta_{\text{PY}}$ satisfies $0 \leq \eta_{\text{PY}} \leq 1$

$$\eta_{\text{PY}} = \sqrt{S(k \to 0)_{\text{HS}}^{\text{PY}} - \sqrt{S(k \to 0)_{\text{HS}}^{\text{PY}} + 3\sqrt{S(k \to 0)_{\text{HS}}^{\text{PY}} + 1}}}.$$

(41)

While Eqs. (18) and (19) are not explicitly concerned with the EHSD, unlike the BH approach, the corresponding EHSD $\sigma_{\text{PY}}$ can still be written as

$$\sigma_{\text{PY}} = \left\{ \frac{6}{\pi \rho} \cdot \left[ \sqrt{S(k \to 0)_{\text{HS}}^{\text{PY}} - \sqrt{S(k \to 0)_{\text{HS}}^{\text{PY}} + 3\sqrt{S(k \to 0)_{\text{HS}}^{\text{PY}} + 1}}} \right] \right\}^{\frac{1}{3}}.$$

(42)

In terms of the EHSD, the fluctuation matching approach provides an EHSD value from a single calculation without any iterations in contrast to the WCA approach.

## 2. Carnahan-Starling EOS

An elegant phenomenological attempt to formulate an accurate, yet empirical, hard sphere EOS was proposed by Carnahan and Starling[99] in order to solve the thermodynamic inconsistency problem of the PY EOS obtained from the virial and compressibility routes:

$$\mathbb{Z}_{\text{CS}} = \frac{1 + \eta + \eta^2 - \eta^3}{(1 - \eta)^3}.$$

(43)

We note that Eq. (43) can be derived by fitting the integer part of the virial coefficient $b_k$ as $k^2 + k - 2$ up to the first six coefficients[118] or equivalently obtained by a simple interpolation form:

$$\mathbb{Z}_{\text{CS}}(\eta) = \frac{1}{3} \mathbb{Z}_{\text{PY}}^v(\eta) + \frac{2}{3} \mathbb{Z}_{\text{PY}}^c(\eta).$$

(44)

The Carnahan-Starling (CS) EOS is almost exact over the stable liquid regime[81] and even significantly into the metastable regime.[119] Thus, we now choose the CS EOS to apply fluctuation matching. From Eq. (18), the partial derivative $\partial P/\partial \eta$ is

$$\left( \frac{\partial P}{\partial \eta} \right) = \frac{\partial}{\partial \eta} \left[ \frac{6k_B T}{\pi \sigma^3} \cdot \frac{\eta + \eta^2 + \eta^3 - \eta^4}{(1 - \eta)^3} \right] = \frac{6k_B T}{\pi \sigma^3} \cdot \left( \frac{\eta^4 - 4\eta^3 + 4\eta^2 + 4\eta + 1}{(1 - \eta)^4} \right).$$

(45)

Utilizing the chain rule and Eq. (45), one has

$$\left( \frac{\partial P}{\partial V} \right) = -\frac{\rho}{V} k_B T \cdot \frac{\eta^4 - 4\eta^3 + 4\eta^2 + 4\eta + 1}{(1 - \eta)^4}.$$

(46)

Finally, the compressibility factor for the CS EOS is then given as

$$S(k = 0)_{\text{HS}}^{\text{CS}} = \frac{(1 - \eta)^4}{\eta^4 - 4\eta^3 + 4\eta^2 + 4\eta + 1}.$$

(47)

### 3. Carnahan-Starling-Kolafa EOS

Even though the CS EOS generally provides an excellent approximate EOS, it can be improved at very high densities while retaining its simple form corresponding to the Carnahan-Starling-Kolafa (CSK) form:[120, 121]

$$\mathbb{Z}_{\text{CSK}} = \frac{1 + \eta + \eta^2 - \frac{2}{3}(\eta^3 + \eta^4)}{(1-\eta)^3}.$$

(48)

Compared to Eq. (43), Equation (48) correctly captures higher-order contributions in density effects. This slight adjustment has been shown to further enhance the evaluation of both the virial coefficients and the compressibility factors.[122]

Other than the aforementioned EOSs, we note that there are numerous empirically or semi-empirically designed EOSs for hard disks, hard spheres, or even hard hyperspheres (e.g., Table 3.11 in Ref. 60 and Table 1 in Ref. 80). However, most of them have complicated polynomial forms that rely on numerical fitting procedures in contrast to the PY and or CS EOSs, which we focus on in this paper. Nevertheless, any of these complicated EOSs could be employed in the proposed framework. From the compressibility factor $\mathbb{Z}_{\text{CSK}}$, the pressure is

$$TP = \frac{6k_B T}{\pi \sigma^3} \cdot \frac{\eta + \eta^2 + \eta^3 - \frac{2}{3}(1+\eta)\cdot \eta^4}{(1-\eta)^3}.$$

(49)

By repeating the procedures employed to obtain Eqs. (45)-(47), we arrive at the following fluctuation matching equation for the CSK EOS:

$$S(k=0)_{\text{HS}}^{\text{CSK}} = \frac{3(1-\eta)^4}{4\eta^5 - 8\eta^4 - 8\eta^3 + 12\eta^2 + 12\eta + 3}.$$

(50)

### 4. CG Fluctuation: Finite Size Effect

In order to solve Eq. (19) in practice, we evaluate the structure factor at zero wave vector $S(k=0)$ using $S(k)_{\text{CG}} = 1 + 4\pi\rho \int_0^\infty (g(R) - 1)\frac{\sin(kR)}{kR} R^2 dR$. From the mapped CG trajectory, we applied the Fast Fourier Transformation of $g(R)$ [123] using the LiquidLib suite.[124] The $g(R)$ functions were sampled with a bin size of 0.02 Å. Alternatively, $S(k=0)$ can be computed using direct numerical integration. In this way, in order to obtain an accurate and numerically stable value, the finite size effect of $S(k=0)_{\text{CG}}$ must be considered by integrating up to the finite distance $R_{\text{cut}}$:

$$S(k=0, R_{\text{cut}}) = 1 + 4\pi \int_0^{R_{\text{cut}}} dR' \cdot R'^2 [g(R') - 1].$$

(51)

Differences between the estimated $S(k = 0, R_{\text{cut}})$ and actual $S(k = 0)$ values can be corrected via the scheme proposed by Salacuse et al.:[125]

$$S(k, R_{\text{cut}}) \approx S_N(k, R_{\text{cut}}) + \frac{S(k=0)}{N}\frac{4}{3}\pi R_{\text{cut}}^3 \cdot \left[\frac{3}{(kR_{\text{cut}})^3} \cdot (\sin kR_{\text{cut}} - kR_{\text{cut}} \cos kR_{\text{cut}})\right], \tag{52}$$

where $S_N(k, R_{\text{cut}})$ is the computed structure factor for a system of $N$ particles with a cutoff $R_{\text{cut}}$. We used the maximum possible value for $R_{\text{cut}}$ as half of the system box length. In the $k = 0$ limit, Equation (52) reduces to

$$S(k=0) \approx \frac{S_N(k=0, R_{\text{cut}})}{1 - \frac{1}{N}\frac{4}{3}\pi \rho R_{\text{cut}}^3}. \tag{53}$$

The computed $S(k = 0)_{\text{CG}}$ values from both methods are in close agreement and for water are: 0.1033 for SPC/E, 0.1051 for SPC/Fw, 0.08814 for TIP4P/2005, and 0.09871 for TIP4P/Ice. Importantly, these values at 300 K and 1 atm conditions are within the reported ranges of 0.06–0.08 from experiments and computer simulations.[126, 127] We attribute the slightly overestimated $S(k = 0)$ values in CG systems to minor differences in the $g(R)$ during the coarse-graining process. Another source of error might be due to the finite size effect in estimating $S(k = 0)$. Since thermodynamic conditions studied here are within the normal regime and not near a critical point, the effect of an overly long correlation length should not be pronounced.[128, 129] As expected, we checked that doubling the system size gives a $S(k = 0)$ value that differs by a very small amount of 0.007 for the SPC/Fw at 300 K condition. A comprehensive computational analysis to determine the $S(k = 0)$ values under various force field and temperature conditions will be systematically pursued in future studies.

## 5. Fluctuation Matching: Results

We now solve the fluctuation matching equation, Eq. (18), for each EOS discussed above by equating the dimensionless compressibility expression to the corrected $S(k = 0)$ values from our computer simulations. Table 1 lists the computed packing fractions for the CG water systems at 300 K. Interestingly, the BH and fluctuation matching approaches yield similar $\eta$ values for different choices of force fields and EOSs. Given the different principles underlying these mapping approaches, this agreement indicates that both methods provide nearly similar hard sphere systems with $\Delta\eta \approx 0.01$ for the same FG system. We also note that the $\eta$ values for both approaches are well below $\eta_f = 0.494$, the volume fraction of freezing,[130] validating the underlying assumption to treat CG particles as equilibrium hard sphere liquids not in the metastable or supercooled regime. The relatively low absolute values of the effective packing fraction near 0.3 may seem surprising for a dense liquid. We believe this is likely a consequence of the relatively large value of $S(k = 0)$ due to the high molecular number density and atypical compressibility due to hydrogen bonding of water. For nonpolar or weakly polar molecular liquids, one expects the mapping would deliver packing fractions substantially larger than 0.3. This will be tested in future work.

> **Table 1:** Effective hard sphere packing fractions ($\eta$) of the hard sphere system mapped from the CG water system using the BH scheme and the fluctuation matching approach with the selected EOSs.

| FG force field | Barker-Henderson | Fluctuation matching | | |
|---|---|---|---|---|
| | | Percus-Yevick | Carnahan-Starling | Carnahan-Starling-Kolafa |
| SPC/E | 0.276 | 0.288 | 0.293 | 0.292 |
| SPC/Fw | 0.283 | 0.286 | 0.291 | 0.290 |
| TIP4P/2005 | 0.288 | 0.308 | 0.313 | 0.312 |
| TIP4P/Ice | 0.285 | 0.294 | 0.299 | 0.298 |

## C. Step 2: Estimating the Hard Sphere Diffusion Coefficient
### 1. Percus-Yevick Equation of State

We now determine an analytical form of the hard sphere diffusion coefficient using different EOSs. For the simplest PY EOS, we arrived at the following expression in Section II-F

$$D_0^{PY} \approx \frac{6^{\frac{1}{3}}}{4} \pi^{\frac{1}{6}} \eta^{\frac{1}{3}} \frac{(1-\eta)^2}{\eta(\eta^2 - 2\eta + 4)} \exp\left[\frac{3(2\eta - \eta^2)}{2(1-\eta)^2}\right]. \tag{54}$$

### 2. Carnahan-Starling Equation of State

The CS EOS has a more accurate contact value $g(R = \sigma)$:

$$g(\sigma) = \frac{\mathbb{Z} - 1}{4\eta} = \frac{1 - \eta/2}{(1-\eta)^3}, \tag{55}$$

and excess entropy:

$$S_{ex} = -\int_0^\eta \frac{\mathbb{Z} - 1}{\eta'} d\eta' = -\eta \frac{(4 - 3\eta)}{(1-\eta)^2}. \tag{56}$$

Substituting Eq. (55) and (56) in Eq. (22) yields an expression for the entropy-free diffusion coefficient from the Dzugutov scaling, $D_{Z,CS}^0$

$$D_{Z,CS}^0 = \frac{\pi}{96} \frac{(1-\eta)^6}{\eta^2(2-\eta)^2} \exp\left[\frac{(4\eta - 3\eta^2)}{(1-\eta)^2}\right]. \tag{57}$$

In Step 3, the hard sphere system ($D_{Z,CS}^0$) is mapped back to the original CG system to assess the entropy-free diffusion coefficient $D_{0,CS}^{HS}$ under the Rosenfeld scaling:

$$D_{0,CS}^{HS} \approx \frac{\pi^{\frac{1}{6}}}{48} \cdot 6^{\frac{4}{3}} \cdot \frac{(1-\eta)^3}{\eta^{\frac{2}{3}}(2-\eta)} \exp\left[\frac{(4\eta - 3\eta^2)}{(1-\eta)^2}\right]. \tag{58}$$

Compared to Eq. (54), differences in $D_0^{HS}$ originate from changes in $g(\sigma)$ that affect collision rates and excess entropy terms from the scaling relationship.

### 3. Carnahan-Starling-Kolafa Equation of State

Similarly, $D_0^{HS}$ from the CSK EOS can be derived, and several structural and thermodynamic properties are required in order to derive $D_0^{HS}$. The contact value of the pair correlation and excess entropy are given by

$$g(\sigma) = \frac{Z-1}{4\eta} = \frac{1 - \frac{\eta}{2} + \frac{\eta^2}{12} - \frac{\eta^3}{6}}{(1-\eta)^3},$$
(59)

$$S_{ex} = -\int_0^\eta \frac{Z-1}{\eta'} d\eta' = -\int_0^\eta \frac{4 - 2\eta' + \frac{\eta'^2}{3} - \frac{2}{3}\eta'^3}{(1-\eta')^3} d\eta' = -\frac{5}{3}\ln(1-\eta) - \eta\frac{4\eta^2 - 33\eta + 34}{6(1-\eta)^2}.$$
(60)

Equation (60) is further confirmed by the results in Ref. 131. Then, substituting Eq. (22) into Eq. (60) yields

$$D_Z^0 = \frac{\pi}{24} \cdot \frac{(1-\eta)^{\frac{23}{3}}}{\eta^2 \left(4 - 2\eta + \frac{1}{3}\eta^2 - \frac{2}{3}\eta^3\right)} \exp\left[\eta\frac{4\eta^2 - 33\eta + 34}{6(1-\eta)^2}\right].$$
(61)

Finally, simple manipulations yield an analytic expression for $D_{0,CSK}^{HS}$:

$$D_{0,CSK}^{HS} \approx \frac{\pi^{\frac{1}{6}}}{96} \cdot 6^{\frac{4}{3}} \cdot \frac{(1-\eta)^{\frac{14}{3}}}{\eta^{\frac{2}{3}}\left(1 - \frac{\eta}{2} + \frac{\eta^2}{12} - \frac{\eta^3}{6}\right)} \exp\left[\eta\frac{4\eta^2 - 33\eta + 34}{6(1-\eta)^2}\right].$$
(62)

### D. Step 3: Diffusion Coefficient

We now evaluate the computed entropy-free diffusion coefficient $D_0^{HS}$ of CG water from the dynamically consistent hard sphere system. Table 2 lists the predicted $\ln(D_0^{HS})$ values from both the BH approach and fluctuation matching scheme based on three EOSs (PY, CS, and CSK). We find that both approaches predict $D_0^{HS}$ values close to the CG reference value of $D_0^{CG} = 0.7047$, confirming the validity of our assumptions. This finding seems rather remarkable in the sense that: (1) adoption of a simple hard sphere model can still be effective for estimating the accelerated CG dynamics, and (2) the matching dimensionless compressibility idea is useful for molecular liquids. Overall, $D_0^{HS}$ values estimated by the BH approach incur an error of 11.1%, whereas fluctuation matching yields values that incur errors within 17%. The modest differences between the two approaches for water may not be representative of the general performance of the two methods for other less complex and non-associated molecular liquids, especially in the supercooled regime, where we expect the effective packing fractions will be significantly larger than $\eta \approx 0.3$ obtained for water. This issue will be explored in a future article.

We also assess the accuracy of the specific EOS chosen in terms of reproducing $D_0$ values. While we find that the relative performance of the EOS in reproducing correct CG dynamics generally follows the accuracy of the EOS itself (BH approach), the relative enhancement is minor (within errors of 2%). This low sensitivity can be understood by the relatively low effective packing fractions of the mapped water system where choice of EOS does not result in major variations.[122, 132, 133] We also note that in this case, the back-mapping approximation is a reasonable assumption, giving $\exp(\alpha^{CG} s_{ex}^{CG}) / \exp(s_{ex}^{HS})$ as 1.1.

**Table 2:** Effective "entropy-free" diffusion coefficients $D_0^{HS}$ from the Rosenfeld scaling is predicted from the effective hard sphere systems mapped from the CG water system using the BH and the fluctuation matching approaches with the selected EOSs.

| FG Force field | Reference value[42] | Barker-Henderson | | | Fluctuation Matching | | |
|---|---|---|---|---|---|---|---|
| | | Percus-Yevick | Carnahan-Starling | Carnahan-Starling-Kolafa | Percus-Yevick | Carnahan-Starling | Carnahan-Starling-Kolafa |
| SPC/E | 0.705 | 0.771 | 0.759 | 0.758 | 0.802 | 0.803 | 0.802 |
| SPC/Fw | | 0.789 | 0.777 | 0.776 | 0.796 | 0.797 | 0.795 |
| TIP4P/2005 | | 0.802 | 0.790 | 0.789 | 0.951 | 0.973 | 0.972 |
| TIP4P/Ice | | 0.792 | 0.780 | 0.780 | 0.855 | 0.862 | 0.860 |

In turn, the results listed in Table 2 confirm the high-fidelity nature of our approach for estimating the $D_0$ value by treating CG dynamics with an effective hard sphere model. Taken one step further, we now compare the full diffusion coefficient $D_{HS}$ by including the excess entropy contributions $s_{ex}^{CG}$ to the CG dynamics, as shown in Fig. 3.

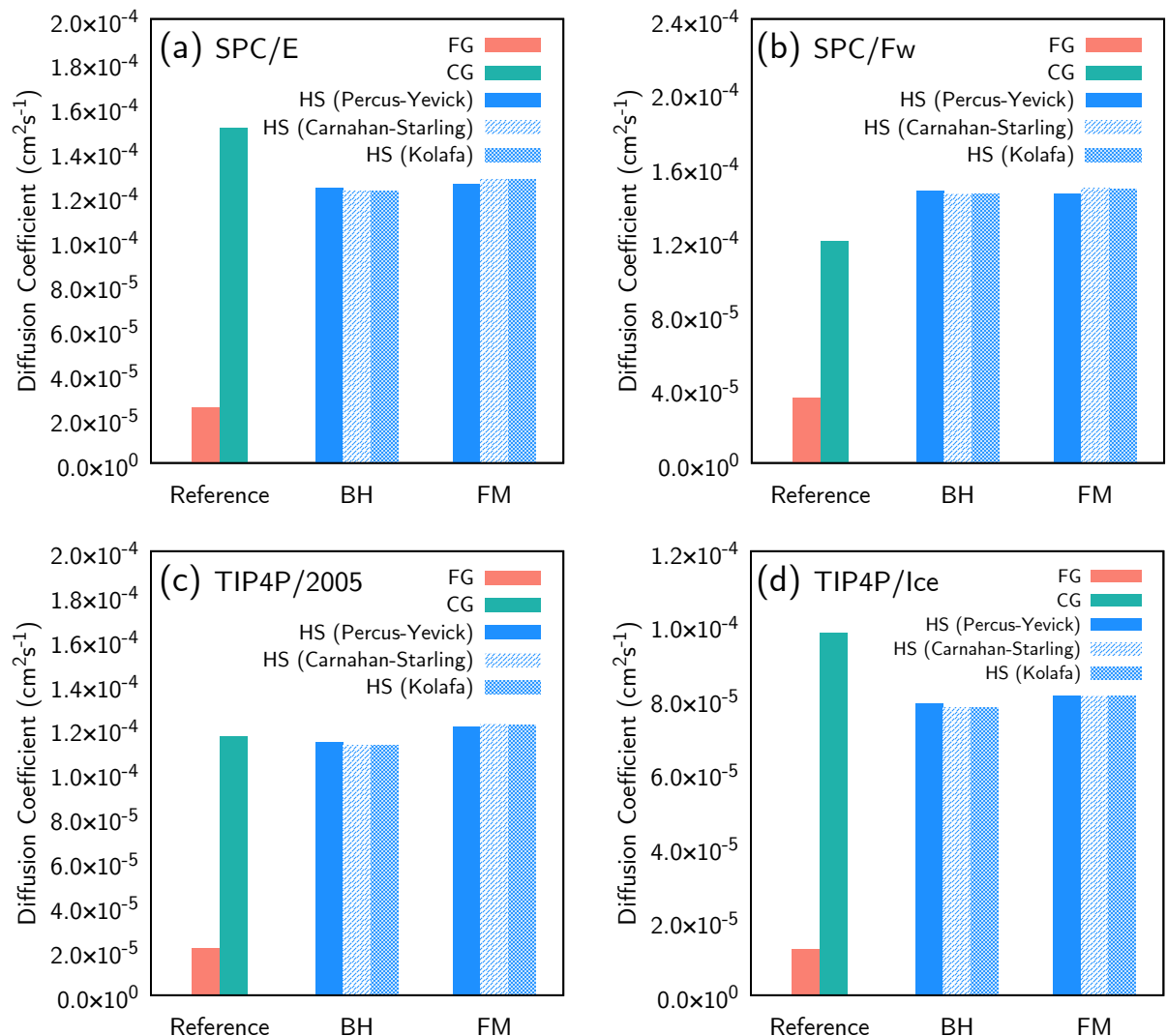

**Figure 3**. Recovered CG diffusion coefficient using the dynamically consistent hard sphere models $D_{HS}$ for four different water force fields at 300 K: (a) SPC/E, (b) SPC/Fw, (d) TIP4P/2005, and (e) TIP4P/Ice. Hard sphere packing fractions or effective diameters are estimated via two different approaches: Barker-Henderson (BH) and fluctuation matching (FM). Then, the corresponding diffusion coefficients are obtained by employing different choices of EOS: Percus-Yevick (blue solid), Carnahan-Starling (half-filled), and Carnahan-Starling-Kolafa (double-filled). References values from FG (red solid) and CG (green solid) systems are included as a comparison.

Figure 3 shows the CG diffusion coefficients recovered from the hard sphere approaches with a correct unit. By comparing these $D_{HS}$ values to the actual numerical CG diffusion coefficients $D_{CG}$, we find, as expected, that both approaches capture well the accelerated CG diffusion coefficients. Interestingly, after incorporating the excess entropy terms, fluctuation matching shows an almost identical level of description (average errors of 15.1%) compared to the BH approach (16.0%). Remarkably, we emphasize that the trend of acceleration factor ascribed to different atomistic force fields is qualitatively captured in our approach. Here, the effective dynamic acceleration factor is estimated by the ratio of $D_{HS}$ to $D_{FG}$. Given the acceleration factors of the CG water models ($D_{CG}/D_{FG}$ is 6.0 for SPC/E, 3.4 for SPC/Fw, 5.5 for TIP4P/2005, and 7.8 for TIP4P/Ice), the hard sphere description correctly captures the accelerated diffusion to be 5.0 for SPC/E, 4.2 for SPC/Fw, 5.6 for TIP4P/2005, and 6.4 for TIP4P/Ice.

**E. Estimation of CG Diffusion Coefficient at Different Temperatures**

We emphasize that the main advantage of fluctuation matching is that we can estimate the overall diffusion coefficient of CG systems *a priori,* solely based on information from the FG systems. This is because if the hard sphere EOS is chosen, the fluctuation matching equation *only* requires $S(k \to 0)$ to match the long wavelength density fluctuations, not the detailed intermolecular potentials. Especially, this feature would be advantageous for some bottom-up CG approaches that aim to reproduce important structural correlations.[10] Among various bottom-up CG methodologies, using pairwise basis sets, the Iterative Boltzmann Inversion (IBI),[134] Inverse Monte Carlo (IMC),[135] and Relative Entropy Minimization (REM)[136, 137] approaches can match the two-body correlations, i.e., RDF. Also, the Multi-Scale Coarse-Graining (MS-CG) approach[9, 109, 138-140] satisfies the Yvon-Born-Green equation in liquid physics,[141] indicating that MS-CG models aim to reproduce up to three-body correlations. Notably, BUMPer is developed upon the MS-CG principle, where the many-body projection theory allows for recapitulating the pairwise correlation as well.[106]

Therefore, for bottom-up CG models designed for capturing structural correlations, i.e., RDF, one can approximate the $S(k \to 0)_{CG}$ without any CG simulation. In other words, with bottom-up approaches, we can approximate the FG RDF as the CG RDF in Eq. (19) to predict $D_0$ values for CG systems at different thermodynamic conditions, especially variable temperature at atmospheric pressure.[142] This estimation is *not possible* in the conventional BH approach, in which the effective CG interactions need to be first parameterized in order to employ Eq. (12).

Based on the predicted $D_0$ from the hard sphere description, a complete determination of $D_{CG}$ is also possible by considering the effective excess entropy contribution remaining at the CG resolution. In Paper I, we elucidated the differences in excess entropy between FG and CG systems corresponding to the mapping entropy, i.e., the missing contributions to the configurational entropies beneath the CG resolution. For single-site CG models, this mapping entropy is the rotational and vibrational entropies from the FG resolution.[79] Therefore, an estimation of CG excess entropy at different temperatures is possible by assessing translational contributions in the FG entropy

$$s_{trn}^{ex}|_{CG} \approx s_{trn}^{FG} - s_{trn}^{(id)} = s_{trn}^{FG} - \left(1.6787 + \frac{3}{2}\ln T\right). \quad (63)$$

where $(1.6787 + 1.5 \ln T)$ is obtained from the translational entropy of ideal gas for this system; please see Ref. 42 for further derivation and analysis. Equation (63) assumes complete entropy representability between the FG and CG systems. However, we adopt an approximate notation instead of an equality since slight differences between $s_{trn}^{ex}|_{FG}$ and $s_{trn}^{ex}|_{CG}$ are expected due to the pairwise approximation introduced in the CG model, slight differences between $s_{trn}^{ex}|_{FG}$ and $s_{trn}^{ex}|_{CG}$ are expected. With this in mind, we compute the estimated CG diffusion coefficients over a range of temperatures from 280 K to 360 K at 20 K intervals by utilizing FG information only. For different temperature conditions, we used the excess entropies and FG diffusion coefficients reported in Paper I.[42]

As depicted in Fig. 4, it is immediately evident that the hard sphere estimation using our approaches successfully recapitulates the CG diffusion coefficients of water over a wide range of temperatures and for different atomistic force fields. For example, in the case of SPC/E, the reference CG diffusion coefficients obtained are: $1.22 \times 10^{-4}$, $1.54 \times 10^{-4}$, $1.77 \times 10^{-4}$, and $1.96 \times 10^{-4}$ cm$^2 \cdot$s$^{-1}$, as temperature increases from 280, 320, 340, and 360 K, respectively. The effect of temperature on the CG diffusion coefficient is well reproduced in the hard sphere model as $9.33 \times 10^{-5}$, $1.68 \times 10^{-4}$, $2.05 \times 10^{-4}$, and $2.55 \times 10^{-4}$ cm$^2 \cdot$s$^{-1}$, respectively. In turn, the hard sphere description provides diffusion coefficients for water similar to the CG model values regardless of the EOS adopted.

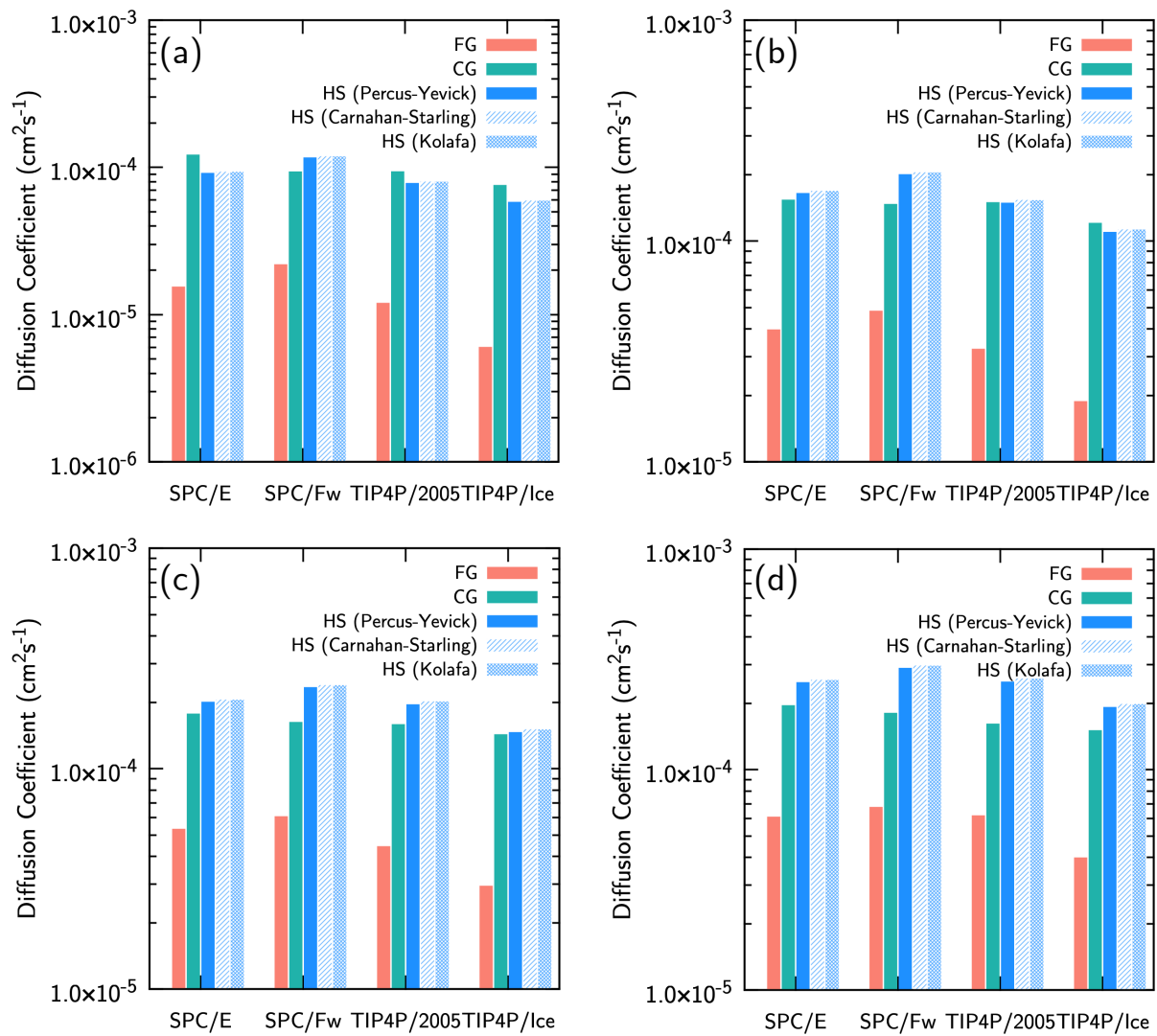

**Figure 4.** CG diffusion coefficients (plotted on a logarithmic scale) predicted from FG information for five different temperatures: (a) 280 K, (b) 320 K, (d) 340 K, and (e) 360 K. The $D_{\text{HS}}$ values are computed by combining the entropy representability relationship from FG information ($s_{\text{trn}}^{ex}|_{\text{FG}}$) and dynamically consistent hard sphere models ($D_0^{\text{HS}}$) using three different EOSs: Percus-Yevick (blue solid), Carnahan-Starling (half-filled), and Carnahan-Starling-Kolafa (double-filled). The predicted hard sphere diffusion

coefficients reasonably reproduce the actual CG values (green solid), which show accelerated dynamics compared to the reference FG (red solid) diffusion coefficients.

We attribute the success of a hard sphere description to the accurate prediction of CG excess entropies from FG models (entropy representability) and to the effective hard sphere nature of normal liquid dynamics, yielding accurate $D_0$ values for different thermodynamic conditions. Although the short-range structural ordering (e.g., RDF intensities at the first peaks) of associated liquids can be significantly affected by temperature, slight changes appear in the dimensionless compressibility, resulting in modest changes of the $\eta$ values. For example, in the case of SPC/E, as $S(k=0)$ varies from 0.10 to 0.088, the resultant $\eta$ values range from 0.28 to 0.31. In other words, from 280 K to 360 K at 1 atm pressure (ambient liquid water condition), we observe that $D_0$ does not change significantly. Such a weak temperature dependence is consistent with the underlying hypothesis of the excess entropy scaling approach that assumes $D_0$ is invariant under changes in temperature and density. Table 3 further supports our claim that the computed $D_0^{HS}$ is not very sensitive to temperature. As discussed earlier, the strongly associated nature of water results in rather large values of $S(k=0)$ and a different temperature dependence compared to nonpolar and weakly polar non-associated liquids.

**Table 3:** Reduced CG diffusion coefficients $D_0^{HS}$ of water predicted from the fluctuation matching approach using the FG compressibilities and the Percus-Yevick, Carnahan-Starling, and Carnahan-Starling-Kolafa EOSs. $D_0^{HS}$ is predicted at temperatures ranging from 280 K to 360 K using four FG force fields: (a) SPC/Fw, (b) SPC/E, (c) TIP4P/2005, and (d) TIP4P/ice.

| (a) SPC/E | | | | (b) SPC/Fw | | | |
|---|---|---|---|---|---|---|---|
| Temp | $D_{0,PY}^{HS}$ | $D_{0,CS}^{HS}$ | $D_{0,CSK}^{HS}$ | Temp | $D_{0,PY}^{HS}$ | $D_{0,CS}^{HS}$ | $D_{0,CSK}^{HS}$ |
| 280 K | 0.778 | 0.796 | 0.794 | 280 K | 0.774 | 0.792 | 0.791 |
| 320 K | 0.827 | 0.849 | 0.848 | 320 K | 0.790 | 0.809 | 0.808 |
| 340 K | 0.826 | 0.849 | 0.847 | 340 K | 0.815 | 0.837 | 0.835 |
| 360 K | 0.847 | 0.872 | 0.871 | 360 K | 0.844 | 0.868 | 0.867 |
| (c) TIP4P/2005 | | | | (d) TIP4P/Ice | | | |
| Temp | $D_{0,PY}^{HS}$ | $D_{0,CS}^{HS}$ | $D_{0,CSK}^{HS}$ | Temp | $D_{0,PY}^{HS}$ | $D_{0,CS}^{HS}$ | $D_{0,CSK}^{HS}$ |
| 280 K | 0.815 | 0.837 | 0.836 | 280 K | 0.799 | 0.820 | 0.818 |
| 320 K | 0.883 | 0.912 | 0.910 | 320 K | 0.863 | 0.889 | 0.888 |
| 340 K | 0.906 | 0.936 | 0.935 | 340 K | 0.892 | 0.921 | 0.919 |
| 360 K | 0.930 | 0.963 | 0.962 | 360 K | 0.918 | 0.950 | 0.948 |

Recently, Bernhardt and Van der Vegt suggested that there may be a single unified excess entropy scaling relationship that encompasses several CG liquid systems.[143] Nevertheless, they also found that the correlation of the putative unified scaling law is not strong and does not agree with the scaling constant of the Rosenfeld relationship.[51-53] Our work resolves these inconsistencies, showing that $D_0$ is quite dependent on the effective packing fraction of different CG molecules, and, thus, a single scaling relationship for many CG systems is not generically valid.

Altogether, our work sheds light on understanding accelerated CG diffusion by focusing on two points: (1) leveraging the entropy representability relationship between FG and CG systems[79] and (2) constructing effective hard sphere systems that approximate CG dynamics. Given the empirical nature of excess entropy scaling relationships, we introduced another layer of coarsening to map CG systems to hard sphere systems where dynamical properties can be formulated analytically. For conditions that the CG particles can be represented as effective hard spheres, we believe that our findings open up a new avenue for bottom-up CG modeling by interpreting CG dynamics from classical perturbation theory and fluctuation matching. As this paper is the first study of such combined efforts, we anticipate several potential directions that can be taken to further analyze the faster CG dynamics for complex molecular systems. One natural extension would be to elucidate the role of resolution in CG dynamics. Since the continuous effort to establish a theoretical link between the CG dynamics and CG resolution has mainly remained *ad hoc*,[144] the proposed methodology is expected to systematically bridge the choice of CG mapping and the resultant dynamics.

## IV. Conclusions

In this paper, we developed a new mapping scheme for molecular liquids in the normal regime from a target coarse-grained (CG) system to an effective hard sphere system as part of our ongoing effort to understand accelerated CG dynamics in terms of excess entropy scaling, which was introduced in Paper I of this series.[42] Even though CG interactions are intrinsically many-body potentials of mean force, the treatment of single-site CG systems as hard spheres is substantiated by perturbation theories of liquids where the repulsive intermolecular interaction primarily determines the structural and dynamical properties. Such simplified models, in which the equation of state (EOS) is a function only of packing fraction, allow for employing analytical theories to predict the diffusion coefficient in CG systems.

In determining the packing fraction, conventional perturbation theories may have a limitation to correctly reproduce dynamics as they mainly aim to determine the effective hard sphere diameter for the purpose of predicting equilibrium structural correlations of a reference fluid that interacts only via harsh repulsive forces from an effective hard sphere model. To construct an effective hard sphere model suitable for describing dynamics in the presence of chemical complexity and attractive interactions, we adopt the idea of Mirigian and Schweizer[72-77] successfully employed for supercooled liquid activated relaxation of molecular and polymeric liquids that the long wavelength amplitude of density fluctuations or dimensionless compressibility is a physically appropriate quantity for constructing a mapping under isobaric conditions in a CG framework. This so-called "fluctuation matching" idea requires the chemistry and thermodynamic state dependent dimensionless compressibilities of the molecular CG and hard sphere systems are exactly equal. The fluctuation matching method directly determines effective hard sphere packing fractions in a manner consistent with the original excess entropy scaling relationship. Along with the conventional BH criterion, we employ fluctuation matching for CG one-site water systems to construct effective hard sphere systems based on adopting different hard sphere fluid EOSs.

Analytical formulations of the entropy-free diffusion coefficient are derived by applying two excess entropy scaling schemes sequentially in an elementary kinetic theory framework for dense liquids. For the mapped hard sphere system, we determine the dynamic properties using the well-known Enskog theory. As the Dzugutov scaling is derived for the hard sphere system, we first apply the Dzugutov scaling to obtain the entropy-free diffusion coefficients $D_Z^0$, and then map it back to the Rosenfeld scaling $D_0$ for the original CG system. This scheme allows for estimating the entropy-free diffusion coefficient for the underlying CG system. Given a wide range of applications of excess entropy scaling, this paper opens a promising direction for future research to better characterize the accelerated dynamics of complex molecules at the reduced level, e.g., confined[145, 146] or active matter[147] systems, especially when comprehensive analysis of collective behaviors at longer timescales is computationally challenging.

The key finding from this work is that the estimated entropy-free diffusion coefficient $D_0^{HS}$ is in remarkable agreement with the actual values from the excess entropy scaling of the CG system $D_0^{CG}$, as well as in agreement with the full diffusion coefficient $D^{CG}$. Notably, and nontrivially, we show that fluctuation matching can be faithfully be applied to molecular liquids with (specific) attractive interactions at the CG level and at temperatures in the normal liquid regime. We believe our work demonstrates the efficacy of the hard sphere mapping to single-site CG systems since the less important degrees of freedom are effectively integrated out, and the resultant CG interactions are spherically symmetric. By employing such a minimalist model, we claim that the acceleration in the CG dynamics can be understood from the FG point of view. We further corroborate our claim by successfully predicting temperature-dependent CG diffusion coefficients *a priori* using only FG information combined with fluctuation matching.

Finally, our findings lead to a systematic rationalization of the acceleration factor, $D_{CG}/D_{FG}$. Under the excess entropy scaling formalism, this acceleration can be understood from two contributing factors. While the first term $\exp(\alpha s_{ex}^{CG})/\exp(\alpha s_{ex}^{FG})$ can be understood from the entropy representability relationship, the second term $D_0^{CG}/D_0^{FG}$ is not relatively clear due to $D_0^{FG}$. Returning to the issue of the correspondence between FG and CG scaling relationships demonstrated in Paper I,[42] a relevant follow-up question based on the present article is how can we physically understand $D_0^{FG}$? Unlike the dynamics at the CG level, dynamics at the full atomistic resolution entail various motions other than pure translation, even at the single-site level of description. However, it might be natural to envision that these other motions, as well as the translational motions, are encoded in the center-of-mass diffusion behavior. In the following paper of this series, we will approach this problem by decoupling rotational motions from translational motion and assessing how rotational diffusion is correlated to the overall translational diffusion. Another possible direction would be to combine fluctuation matching for $S(k \to 0)_{CG}$ with non-hard sphere EOSs that can capture the FG dynamics. Using idealized or semi-empirical EOSs, it is possible that rotational and vibrational contributions deviated from hard sphere translations could be addressed with additional variables. Altogether, these proposed directions may lead to a unified framework for unraveling the fundamental differences underlying FG and CG dynamics.

**ACKNOWLEDGMENTS**
This material is based upon work supported by the National Science Foundation (NSF Grant CHE-2102677). Simulations were performed using computing resources provided by the University of Chicago Research Computing Center (RCC). J.J. acknowledges the Harper Dissertation Fellowship from the University of Chicago and insightful discussions with Professor Jeppe C. Dyre and Professor Eok Kyun Lee. K.S.S acknowledges the support of the University of Chicago and the Pritzker School of Molecular Engineering during his sabbatical stay where this work was initiated.

## DATA AVAILABILITY
The data that support the findings of this work are available from the corresponding author upon request.

## APPENDIX
### A. Percus-Yevick (Virial Route): Fluctuation matching
As introduced in Section II-F, the compressibility factor for the virial route of the PY EOS is[101-103]

$$\mathbb{Z}_{PY}^v = \frac{(1 + 2\eta + 3\eta^2)}{(1-\eta)^2}. \tag{A1}$$

Repeating the analysis employed in Eqs. (38)-(40), we calculate the $\left(\frac{\partial P}{\partial V}\right)$ term using the chain rule

$$\left(\frac{\partial P}{\partial V}\right) = -\frac{6}{\pi\sigma^3} \cdot \left[-\frac{3\eta^3 - 9\eta^2 - 5\eta - 1}{(1-\eta)^3}\right] \cdot \frac{\eta}{V} = -\frac{\rho}{\beta V} \cdot \frac{-3\eta^3 + 9\eta^2 + 5\eta + 1}{(1-\eta)^3}. \tag{A2}$$

Finally, the compressibility factor from the structure factor is obtained as

$$S(k=0)_{HS}^{PY} = \rho k_B T \cdot \left(-\frac{1}{V} \cdot \frac{\partial V}{\partial P}\right) = \frac{(1-\eta)^3}{-3\eta^3 + 9\eta^2 + 5\eta + 1}. \tag{A3}$$

### B. Percus-Yevick (Virial Route): Diffusion Coefficient
As demonstrated in Section II-F, the hard sphere diffusion coefficient is described by Eq. (26), where its complete form can be analytically determined using the EOS or compressibility factor $\mathbb{Z}$. In this section, we derive an exact expression of $D_0$ using the virial route PY EOS. First, we calculate the contact value of the radial distribution function:

$$g(\sigma) = \frac{1 + \frac{\eta}{2}}{(1-\eta)^2}. \tag{B1}$$

The excess entropy can be expressed as

$$S_{ex}^{HS} = -\int_0^\eta \frac{\mathbb{Z}-1}{\eta'} d\eta' = -\int_0^\eta 2\frac{\eta'+2}{(1-\eta')^2} d\eta' = -2\ln(1-\eta) - \frac{6\eta}{1-\eta}. \tag{B2}$$

Combining Eqs. (B1) and (B2), we obtain

$$D_Z^0 = \frac{\pi}{96} \cdot \frac{(1-\eta)^6}{\eta^2(2+\eta)^2} \exp\left(\frac{6\eta}{1-\eta}\right).$$

(B3)

Therefore, the full diffusion coefficient is expressed as

$$D = \frac{\pi^2}{48} \sigma^4 \rho \sqrt{\frac{k_B T}{m}} \cdot \left[\frac{(1-\eta)^4}{\eta^2(2+\eta)}\right] \exp\left(\frac{6\eta}{1-\eta}\right) \exp(s_{ex}).$$

(B4)

After mapping $D$ in Eq. (B4) back to the CG system, we obtain

$$D_0^{\text{PY}} \approx \frac{6^{\frac{1}{3}}}{8} \pi^{\frac{1}{6}} \eta^{\frac{1}{3}} \frac{(1-\eta)^4}{\eta(2+\eta)} \exp\left(\frac{6\eta}{1-\eta}\right).$$

(B5)

**REFERENCES**
1. F. Müller-Plathe, ChemPhysChem **3** (9), 754-769 (2002).
2. H. A. Scheraga, M. Khalili and A. Liwo, Annu. Rev. Phys. Chem. **58**, 57-83 (2007).
3. G. A. Voth, *Coarse-graining of condensed phase and biomolecular systems*. (CRC press, 2008).
4. C. Peter and K. Kremer, Soft Matter **5** (22), 4357-4366 (2009).
5. T. Murtola, A. Bunker, I. Vattulainen, M. Deserno and M. Karttunen, Phys. Chem. Chem. Phys. **11** (12), 1869-1892 (2009).
6. S. Riniker and W. F. van Gunsteren, J. Chem. Phys. **134** (8), 084110 (2011).
7. M. G. Saunders and G. A. Voth, Annu. Rev. Biophys. **42**, 73-93 (2013).
8. W. G. Noid, J. Chem. Phys. **139** (9), 090901 (2013).
9. L. Lu, S. Izvekov, A. Das, H. C. Andersen and G. A. Voth, J. Chem. Theory Comput. **6** (3), 954-965 (2010).
10. J. Jin, A. J. Pak, A. E. P. Durumeric, T. D. Loose and G. A. Voth, J. Chem. Theory Comput. **18** (10), 5759-5791 (2022).
11. R. Henderson, Phys. Lett. A **49** (3), 197-198 (1974).
12. W. G. Noid, Methods Mol. Biol. (924), 487-531 (2013).
13. J. W. Wagner, J. F. Dama, A. E. P. Durumeric and G. A. Voth, J. Chem. Phys. **145** (4), 044108 (2016).
14. J. Wang, S. Olsson, C. Wehmeyer, A. Pérez, N. E. Charron, G. De Fabritiis, F. Noé and C. Clementi, ACS Cent. Sci. **5** (5), 755-767 (2019).
15. T. Kinjo and S.-a. Hyodo, Phys. Rev. E **75** (5), 051109 (2007).
16. C. Hijón, P. Español, E. Vanden-Eijnden and R. Delgado-Buscalioni, Faraday Discuss. **144**, 301-322 (2010).
17. H. Lei, B. Caswell and G. E. Karniadakis, Phys. Rev. E **81** (2), 026704 (2010).
18. L. Gao and W. Fang, J. Chem. Phys. **135** (18), 184101 (2011).
19. S. Izvekov, J. Chem. Phys. **138** (13), 134106 (2013).
20. Y. Yoshimoto, I. Kinefuchi, T. Mima, A. Fukushima, T. Tokumasu and S. Takagi, Phys. Rev. E **88** (4), 043305 (2013).
21. Z. Li, X. Bian, B. Caswell and G. E. Karniadakis, Soft Matter **10** (43), 8659-8672 (2014).

22. A. Davtyan, J. F. Dama, G. A. Voth and H. C. Andersen, J. Chem. Phys. **142** (15), 154104 (2015).
23. A. Davtyan, G. A. Voth and H. C. Andersen, J. Chem. Phys. **145** (22), 224107 (2016).
24. Z. Li, X. Bian, X. Li and G. E. Karniadakis, J. Chem. Phys. **143** (24), 243128 (2015).
25. Z. Li, H. S. Lee, E. Darve and G. E. Karniadakis, J. Chem. Phys. **146** (1), 014104 (2017).
26. H. Lei, X. Yang, Z. Li and G. E. Karniadakis, J. Comput. Phys. **330**, 571-593 (2017).
27. S. Izvekov, J. Chem. Phys. **146** (12), 124109 (2017).
28. S. Izvekov, Phys. Rev. E **95** (1), 013303 (2017).
29. G. Jung, M. Hanke and F. Schmid, J. Chem. Theory Comput. **13** (6), 2481-2488 (2017).
30. G. Jung, M. Hanke and F. Schmid, Soft Matter **14** (46), 9368-9382 (2018).
31. N. Bockius, J. Shea, G. Jung, F. Schmid and M. Hanke, J. Phys.: Condens. Matter **33** (21), 214003 (2021).
32. F. Glatzel and T. Schilling, Europhys. Lett. **136** (3), 36001 (2021).
33. Y. Han, J. Jin and G. A. Voth, J. Chem. Phys. **154** (8), 084122 (2021).
34. P. K. Depa and J. K. Maranas, J. Chem. Phys. **123** (9), 094901 (2005).
35. V. A. Harmandaris and K. Kremer, Soft Matter **5** (20), 3920-3926 (2009).
36. V. A. Harmandaris and K. Kremer, Macromolecules **42** (3), 791-802 (2009).
37. D. Fritz, K. Koschke, V. A. Harmandaris, N. F. van der Vegt and K. Kremer, Phys. Chem. Chem. Phys. **13** (22), 10412-10420 (2011).
38. P. Depa, C. Chen and J. K. Maranas, J. Chem. Phys. **134** (1), 014903 (2011).
39. K. M. Salerno, A. Agrawal, D. Perahia and G. S. Grest, Phys. Rev. Lett. **116** (5), 058302 (2016).
40. J. F. Rudzinski, Computation **7** (3), 42 (2019).
41. V. Klippenstein, M. Tripathy, G. Jung, F. Schmid and N. F. van der Vegt, J. Phys. Chem. B **125**, 4931-4954 (2021).
42. J. Jin, K. S. Schweizer and G. A. Voth, arXiv preprint arXiv:2208.00078 (2022).
43. R. Zwanzig, J. Chem. Phys. **33** (5), 1338-1341 (1960).
44. R. Zwanzig, Phys. Rev. **124** (4), 983 (1961).
45. R. Zwanzig, Physica **30** (6), 1109-1123 (1964).
46. H. Mori, Prog. Theor. Phys. **33** (3), 423-455 (1965).
47. I. Lyubimov, J. McCarty, A. Clark and M. Guenza, J. Chem. Phys. **132** (22), 224903 (2010).
48. I. Lyubimov and M. Guenza, Phys. Rev. E **84** (3), 031801 (2011).
49. I. Lyubimov and M. Guenza, J. Chem. Phys. **138** (12), 12A546 (2013).
50. P. Stinis, Proc. R. Soc. London, Ser. A **471** (2176), 20140446 (2015).
51. Y. Rosenfeld, Phys. Rev. A **15** (6), 2545 (1977).
52. Y. Rosenfeld, Chem. Phys. Lett. **48** (3), 467-468 (1977).
53. Y. Rosenfeld, J. Phys.: Condens. Matter **11** (28), 5415 (1999).
54. M. Dzugutov, Nature (London) **381** (6578), 137-139 (1996).
55. A. Samanta, S. M. Ali and S. K. Ghosh, Phys. Rev. Lett. **87** (24), 245901 (2001).
56. A. Samanta, S. M. Ali and S. K. Ghosh, Phys. Rev. Lett. **92** (14), 145901 (2004).
57. M. K. Nandi, A. Banerjee, S. Sengupta, S. Sastry and S. M. Bhattacharyya, J. Chem. Phys. **143** (17), 174504 (2015).
58. S. Acharya and B. Bagchi, J. Chem. Phys. **153**, 184701 (2020).
59. J.-P. Hansen and I. R. McDonald, *Theory of simple liquids*. (Elsevier, 1990).

60. Á. Mulero, *Theory and simulation of hard-sphere fluids and related systems*. (Springer, 2008).
61. A. Santos, *A Concise Course on the Theory of Classical Liquids*. (Springer, Berlin, 2016).
62. J. V. Sengers, R. Kayser, C. Peters and H. White, *Equations of state for fluids and fluid mixtures*. (Elsevier, 2000).
63. R. W. Zwanzig, J. Chem. Phys. **22** (8), 1420-1426 (1954).
64. J. A. Barker and D. Henderson, J. Chem. Phys. **47** (8), 2856-2861 (1967).
65. J. A. Barker and D. Henderson, J. Chem. Phys. **47** (11), 4714-4721 (1967).
66. J. D. Weeks, D. Chandler and H. C. Andersen, J. Chem. Phys. **54** (12), 5237-5247 (1971).
67. H. C. Andersen, J. D. Weeks and D. Chandler, Phys. Rev. A **4** (4), 1597 (1971).
68. J. D. Weeks, D. Chandler and H. C. Andersen, J. Chem. Phys. **55** (11), 5422-5423 (1971).
69. B. Widom, Science **157** (3787), 375-382 (1967).
70. L. Verlet, Phys. Rev. **159** (1), 98 (1967).
71. L. Verlet and J.-J. Weis, Phys. Rev. A **5** (2), 939 (1972).
72. S. Mirigian and K. S. Schweizer, J. Phys. Chem. Lett. **4** (21), 3648-3653 (2013).
73. S. Mirigian and K. S. Schweizer, J. Chem. Phys. **140** (19), 194506 (2014).
74. S. Mirigian and K. S. Schweizer, J. Chem. Phys. **140** (19), 194507 (2014).
75. S. Mirigian and K. S. Schweizer, Macromolecules **48** (6), 1901-1913 (2015).
76. B. Mei, Y. Zhou and K. S. Schweizer, Proc. Natl. Acad. Sci. U.S.A. **118** (18), e2025341118 (2021).
77. B. Mei, Y. Zhou and K. S. Schweizer, Macromolecules **54** (21), 10086-10099 (2021).
78. M. S. Shell, J. Chem. Phys. **137** (8), 084503 (2012).
79. J. Jin, A. J. Pak and G. A. Voth, J. Phys. Chem. Lett. **10** (16), 4549-4557 (2019).
80. A. Mulero, C. Galan and F. Cuadros, Phys. Chem. Chem. Phys. **3** (22), 4991-4999 (2001).
81. J. Kolafa, S. Labík and A. Malijevský, Phys. Chem. Chem. Phys. **6** (9), 2335-2340 (2004).
82. J. Kolafa, Phys. Chem. Chem. Phys. **8** (4), 464-468 (2006).
83. J. M. Trusler, in *Applied Thermodynamics of Fluids* (RSC Publishing Cambridge, UK, 2010), pp. 33-52.
84. A. D. Phan, J. Knapik-Kowalczuk, M. Paluch, T. X. Hoang and K. Wakabayashi, Mol. Pharmaceutics **16** (7), 2992-2998 (2019).
85. A. D. Phan, A. Jedrzejowska, M. Paluch and K. Wakabayashi, ACS Omega **5** (19), 11035-11042 (2020).
86. B. Mei, Y. Zhou and K. S. Schweizer, J. Phys. Chem. B **124** (28), 6121-6131 (2020).
87. Y. Zhou, B. Mei and K. S. Schweizer, Phys. Rev. E **101** (4), 042121 (2020).
88. A. Ghanekarade, A. D. Phan, K. S. Schweizer and D. S. Simmons, Proc. Natl. Acad. Sci. U.S.A. **118** (31) (2021).
89. J.-L. Bretonnet, J. Chem. Phys. **117** (20), 9370-9373 (2002).
90. N. Jakse and A. Pasturel, Sci. Rep. **6**, 20689 (2016).
91. D. Enskog, Kungl. Svenska Vetenskap. Handl **63** (1-44), 42 (1922).

92. S. Chapman, T. G. Cowling and D. Burnett, *The mathematical theory of non-uniform gases: an account of the kinetic theory of viscosity, thermal conduction and diffusion in gases*. (Cambridge university press, 1990).
93. H. Longuet-Higgins and J. Pople, J. Chem. Phys. **25** (5), 884-889 (1956).
94. B. Alder, D. Gass and T. Wainwright, J. Chem. Phys. **53** (10), 3813-3826 (1970).
95. J. J. Erpenbeck and W. W. Wood, Phys. Rev. A **43** (8), 4254 (1991).
96. J. K. Percus and G. J. Yevick, Phys. Rev. **110** (1), 1 (1958).
97. B. Widom, J. Chem. Phys. **39** (11), 2808-2812 (1963).
98. H. Reiss, H. Frisch and J. Lebowitz, J. Chem. Phys. **31** (2), 369-380 (1959).
99. M. Mandell and H. Reiss, J. Stat. Phys. **13** (2), 113-128 (1975).
100. M. Heying and D. S. Corti, J. Phys. Chem. B **108** (51), 19756-19768 (2004).
101. E. Thiele, J. Chem. Phys. **39** (2), 474-479 (1963).
102. M. Wertheim, Phys. Rev. Lett. **10** (8), 321 (1963).
103. M. Wertheim, J. Math. Phys. **5** (5), 643-651 (1964).
104. W. Wang, M. K. Khoshkbarchi and J. H. Vera, Fluid Phase Equilib. **115** (1-2), 25-38 (1996).
105. M. Baus and J.-L. Colot, Phys. Rev. A **36** (8), 3912 (1987).
106. J. Jin, Y. Han, A. J. Pak and G. A. Voth, J. Chem. Phys. **154** (4), 044104 (2021).
107. J. Jin, A. J. Pak, Y. Han and G. A. Voth, J. Chem. Phys. **154** (4), 044105 (2021).
108. V. Molinero and E. B. Moore, J. Phys. Chem. B **113** (13), 4008-4016 (2008).
109. S. Izvekov and G. A. Voth, J. Chem. Phys. **123** (13), 134105 (2005).
110. L. Larini, L. Lu and G. A. Voth, J. Chem. Phys. **132** (16), 164107 (2010).
111. F. H. Stillinger and T. A. Weber, Phys. Rev. B **31** (8), 5262 (1985).
112. J. Hoyt, M. Asta and B. Sadigh, Phys. Rev. Lett. **85** (3), 594 (2000).
113. G. Li, C. Liu and Z. Zhu, J. Non-Cryst. Solids **351** (10-11), 946-950 (2005).
114. Y. D. Fomin, V. Ryzhov and N. Gribova, Phys. Rev. E **81** (6), 061201 (2010).
115. D. Chandler, Acc. Chem. Res. **7** (8), 246-251 (1974).
116. R. A. Horne, *Water and aqueous solutions: structure, thermodynamics, and transport processes*. (John Wiley & Sons, 1971).
117. W. R. Smith and I. Nezbeda, J. Chem. Phys. **81** (8), 3694-3699 (1984).
118. N. Clisby and B. M. McCoy, J. Stat. Phys. **122** (1), 15-57 (2006).
119. L. Fernández, V. Martín-Mayor, B. Seoane and P. Verrocchio, Phys. Rev. Lett. **108** (16), 165701 (2012).
120. T. Boublík, Mol. Phys. **59** (2), 371-380 (1986).
121. T. Boublík and I. Nezbeda, Collect. Czech. Chem. Commun. **51** (11), 2301-2432 (1986).
122. J. J. Erpenbeck and W. W. Wood, J. Stat. Phys. **35** (3-4), 321-340 (1984).
123. K. Zhang, arXiv preprint arXiv:1606.03610 (2016).
124. N. P. Walter, A. Jaiswal, Z. Cai and Y. Zhang, Comput. Phys. Commun. **228**, 209-218 (2018).
125. J. Salacuse, A. Denton and P. Egelstaff, Phys. Rev. E **53** (3), 2382 (1996).
126. G. N. Clark, G. L. Hura, J. Teixeira, A. K. Soper and T. Head-Gordon, Proc. Natl. Acad. Sci. U.S.A. **107** (32), 14003-14007 (2010).
127. S. Overduin and G. Patey, J. Phys. Chem. B **116** (39), 12014-12020 (2012).
128. S. Overduin and G. Patey, J. Chem. Phys. **143** (9), 094504 (2015).
129. J. Guo, R. S. Singh and J. C. Palmer, Mol. Phys. **116** (15-16), 1953-1964 (2018).
130. M. Rintoul and S. Torquato, Phys. Rev. Lett. **77** (20), 4198 (1996).
33

131. D. M. Heyes, M. Cass, J. G. Powles and W. Evans, J. Phys. Chem. B **111** (6), 1455-1464 (2007).
132. R. Jadrich and K. S. Schweizer, J. Chem. Phys. **139** (5), 054501 (2013).
133. R. Jadrich and K. S. Schweizer, J. Chem. Phys. **139** (5), 054502 (2013).
134. D. Reith, M. Pütz and F. Müller-Plathe, J. Comput. Chem. **24** (13), 1624-1636 (2003).
135. A. P. Lyubartsev and A. Laaksonen, Phys. Rev. E **52** (4), 3730 (1995).
136. M. S. Shell, J. Chem. Phys. **129** (14), 144108 (2008).
137. A. Chaimovich and M. S. Shell, Phys. Rev. E **81** (6), 060104 (2010).
138. S. Izvekov and G. A. Voth, J. Phys. Chem. B **109** (7), 2469-2473 (2005).
139. W. G. Noid, J.-W. Chu, G. S. Ayton, V. Krishna, S. Izvekov, G. A. Voth, A. Das and H. C. Andersen, J. Chem. Phys. **128** (24), 244114 (2008).
140. W. G. Noid, P. Liu, Y. Wang, J.-W. Chu, G. S. Ayton, S. Izvekov, H. C. Andersen and G. A. Voth, J. Chem. Phys. **128** (24), 244115 (2008).
141. W. G. Noid, J.-W. Chu, G. S. Ayton and G. A. Voth, J. Phys. Chem. B **111** (16), 4116-4127 (2007).
142. J. Jin, A. Yu and G. A. Voth, J. Chem. Theory Comput. **16** (11), 6823-6842 (2020).
143. M. P. Bernhardt, M. Dallavalle and N. F. Van der Vegt, Soft Mater. **0** (0), 1-16 (2020).
144. G. G. Rondina, M. C. Böhm and F. Müller-Plathe, J. Chem. Theory Comput. **16** (3), 1431-1447 (2020).
145. J. Mittal, J. R. Errington and T. M. Truskett, Phys. Rev. Lett. **96** (17), 177804 (2006).
146. J. Mittal, J. R. Errington and T. M. Truskett, J. Phys. Chem. B **111** (34), 10054-10063 (2007).
147. S. A. Ghaffarizadeh and G. J. Wang, J. Phys. Chem. Lett. **13**, 4949-4954 (2022).